\documentclass[aps,prx,preprint,onecolumn]{revtex4-1}  
\usepackage{amsmath,amssymb,bm,bbm} 
\usepackage{graphicx}  
\usepackage{color} 
\usepackage[dvipsnames]{xcolor}
\usepackage[papersize={8.5in,11in}]{geometry}
\usepackage[colorlinks=true]{hyperref}  
\usepackage{ulem}
\usepackage[mathscr]{euscript}
\usepackage[section]{placeins}
\hypersetup{
    unicode=false,          
    pdftoolbar=true,        
    pdfmenubar=true,        
    pdffitwindow=false,     
    pdfstartview={FitH},    
    pdfsubject={},   
    pdftitle={Bilocal quantum criticality},    
    pdfauthor={Scammell et al},     
    pdfcreator={},   
    pdfproducer={}, 
    pdfkeywords={} {} {}, 
    pdfnewwindow=true,      
    colorlinks=true,       
    linkcolor=magenta, 
    citecolor=blue,        
    filecolor=magenta,      
    urlcolor=blue           
 }

\usepackage{cleveref}

\usepackage{amsfonts}
\usepackage{upgreek}
\usepackage{slashed}
\usepackage{latexsym}

\newcommand{\beq}{\begin{equation}}
\newcommand{\eeq}{\end{equation}}
\def\bea{\begin{eqnarray}}
\def\eea{\end{eqnarray}}

\newcommand{\nn}{\nonumber \\}

\usepackage{braket}
\usepackage{comment}
\usepackage{slashed}
\usepackage{subcaption}

\begin{document}


\title{Bilocal quantum criticality}

\author{Harley D. Scammell}
\affiliation{Department of Physics, Harvard University, Cambridge, MA 02138, USA}

\author{Mathias S. Scheurer}
\affiliation{Department of Physics, Harvard University, Cambridge, MA 02138, USA}

\author{Subir Sachdev}
\affiliation{Department of Physics, Harvard University, Cambridge, MA 02138, USA}

\date{\today
\\
\vspace{0.4in}}

\begin{abstract}
We consider 2+1 dimensional conformal gauge theories coupled to additional degrees of freedom which induce a spatially local but long-range in time $1/(\tau-\tau')^2$ interaction
between gauge-neutral local operators. Such theories have been argued to describe the hole-doped cuprates near optimal doping. We focus on a SU(2) gauge theory with $N_h$ flavors of adjoint Higgs fields undergoing a quantum transition between Higgs and confining phases: the $1/(\tau-\tau')^2$ interaction arises from a spectator large Fermi surface of electrons. The large $N_h$ expansion leads to an effective action containing fields which are bilocal in time but local in space. We find a strongly-coupled fixed point at order $1/N_h$, with dynamic critical exponent $z > 1$. We show that the entropy preserves hyperscaling, but nevertheless leads to a linear in temperature specific heat with a co-efficient which has a finite enhancement near the quantum critical point.
\end{abstract}

\maketitle
\tableofcontents

\section{Introduction}
\label{sec:intro}

Strongly-coupled gauge theories in 2+1 spacetime dimensions play a fundamental role in many phenomena in quantum condensed matter physics. Of special interest are `deconfined' critical points of such theories, which separate phases with different patterns of confinement, broken symmetry, and/or topological order. The best understood class of such critical points have an emergent relativistic conformal symmetry, allowing use of many tools from the conformal field theory literature. However, such conformal gauge theories apply to limited classes of phenomena in insulators or quantum Hall systems, and usually not to metallic, compressible systems with Fermi surfaces in the clean limit. 
In particular, conformal critical systems have a low temperature ($T$) specific heat $C_v \sim T^2$, which is smaller than the specific heat $C_v \sim T$ in metals.

Our interest in studying gauge theories of critical points in metals was motivated by
numerous experimental indications \cite{CPLT18,Vishik2012,He14,Fujita14,Badoux16,Loram01,Michon18,Loram19,Bourges18,Shen19,CPana1,CPana2,Julien19} of optimal doping criticality in the hole-doped cuprate superconductors.
We examine here further aspects of a recently proposed \cite{SSST19,SPSS20} SU(2) gauge theory for the vicinity of optimal doping in which a parent conformal theory is coupled to a large Fermi surface of gauge-neutral electrons. 
This theory describes a phase transition from a Higgs phase, representing the pseudogap regime, to a confining phase, representing the overdoped Fermi liquid.
The main effect of the spectator Fermi surface is 
a spatially local, but long-range in time interaction $\sim 1/(\tau-\tau')^2$ between gauge-neutral local operators, where $\tau,\tau'$ are the imaginary time co-ordinates of two such operators \cite{Morinari02,Zaanen02A,Zaanen02B,TGTS10,Kaul08,Mross12,Mross12b}. Within the context of a $1/N_h$ expansion, where $N_h$ is the number of flavors of Higgs fields, we find that this long-range interaction leads to a field theory that is {\it bilocal\/} in time, but local in space {\it i.e.\/} some fields depend upon one spatial co-ordinate $x$, and two time co-ordinates $\tau$ and $\tau'$. 

The bilocality is a consequence of the spectator Fermi surface. In the vicinity of conventional symmetry breaking transitions, Hertz \cite{hertz} argued that the low energy excitations on the Fermi surface could be accounted for by long-range interactions between the order parameter fields. Such arguments were extended to the SU(2) gauge theory in Ref.~\onlinecite{SSST19}, and in the case of interest to us, the long-range interactions induced by the large Fermi surface were irrelevant near the upper critical spatial dimension $d=3$. However, as we will describe in detail here, the long-range interactions are relevant for the SU(2) gauge theory in $d=2$ in the large $N_h$ limit, and lead to a bilocal field theory for computing the $1/N_h$ expansion.

We will find that the bilocal criticality is described by
a new fixed point. This new fixed point is not relativistically invariant, and has dynamic critical exponent $z > 1$ (see (\ref{zres})). We show that the free energy preserves hyperscaling {\it i.e\/} its leading singular term is consistent with scaling dimension $d+z$. At first sight, this suggests that there is no contribution to a linear in $T$ specific heat from the singular hyperscaling preserving term. This turns out to not be the case. The specific heat is given by
\beq
C_v =\gamma_b \, T  + T^{d/z} \, \Phi \left(\frac{T}{\Delta}\right)\,. \label{Cscale}
\eeq
Here $\gamma_b T$ is the background and non-critical specific heat from the spectator Fermi surface;
the prefactor $\gamma_b$ evolves smoothly across the critical point.
The second term is the interesting, singular hyperscaling preserving term, with $\Phi$ a scaling function, and $\Delta$ an energy scale measuring the distance from the quantum critical point on the confining/Fermi liquid side. In the Fermi liquid regime, $T \ll \Delta$, this hyperscaling preserving term contributes $C_v/T \sim - \Delta^{d/z-1}$, and so yields a finite enhancement of $\lim_{T \rightarrow 0}C_v/T$ near the quantum critical point.

We note that bilocal field theories have appeared earlier in the context of systems with random interactions. Ref.~\onlinecite{RSY95} obtained a bilocal field theory for the Ising spin glass in a transverse field. Bilocal field theories also play a central role in models with random and all-to-all interactions, and in particular those related to the Sachdev-Ye-Kitaev (SYK) models \cite{GPS00,GPS01,Maldacena2016,KitaevSuh,Gu2019}. Our model appears to be the first realization in a non-random system, and the bilocality arises from a subtle interplay between the gauge-charged matter fields, and the Fermi surface of electrons. Given the phenomenological appeal of SYK models, the appearance of bilocality in more realistic models of cuprate physics is encouraging.

We will introduce our model field theory with bilocality in Section~\ref{sec:model}. We describe $N_h = \infty$ saddle-point theory in Section~\ref{sec:saddle}. A key ingredient in our analysis is the bilocal field $C (x,\tau,\tau')$, and we describe its low $T$ saddle point value $C(\tau - \tau')$ in Sections~\ref{sec:Crit}-\ref{sec:nonzeroT}. We will compute the free energy at $N_h=\infty$ in Section~\ref{sec:freeen}. Full numerical solutions of the saddle point equations appears in Section~\ref{sec:numerics}. We turn to a renormalization group analysis in Section~\ref{sec:rg}, where we will obtain some results to order $1/N_h$, including the value of $z$ in (\ref{zres}).

\section{The Model}
\label{sec:model}

The gauge-charged matter sector of the model of Refs.~\onlinecite{SSST19,SPSS20} has real Higgs fields $H_{a\ell}$, where $a=1,2,3$ is the SU(2) adjoint gauge index, and $\ell = 1 \ldots N_h$ is the flavor index. This is coupled to SU(2) gauge field $A_{a\mu}$, where $\mu$ is a spacetime index. 
The Higgs field arises from a  transformation of the spin density wave order parameter to a rotating reference frame, and optimal doping criticality is mapped onto the Higgs-confinement transition of such a gauge theory.
The continuum action for the theory is $\int d^d x d \tau \mathcal{L}_H + \mathcal{S}_f$ (we set $d=2$)
with the Lagrangian density
\beq
\mathcal{L}_H = \frac{1}{4g_a^2} F_{a \mu\nu} F_{a \mu \nu} + \frac{1}{2} \left( \partial_\mu H_{a\ell} - \varepsilon_{abc} A_{b\mu}
H_{c\ell} \right)^2 + V(H) \label{LH2}
\eeq
with the field strength
\beq 
F_{a \mu\nu} = \partial_\mu A_{a \nu} - \partial_\nu A_{a \mu} - \varepsilon_{abc} A_{b\mu} A_{c \nu}\,,
\eeq
and the Higgs potential
\begin{align}
\label{HiggsPotential2}
V(H) 
&=  \frac{{u}_0}{2 N_h} \left[H_{a \ell} H_{a \ell} - \frac{3N_h}{g}\right]^2
+ \frac{{u}_1}{2 N_h} H_{a \ell} H_{a m} H_{b \ell} H_{b m} 
\end{align}
The coupling $g$ is the tuning parameter across the Higgs transition. For $g<g_c$, we have the Higgs phase: this is proposed to describe the underdoped pseudogap regime of the cuprates, and its properties were discussed in detail in Ref.~\onlinecite{SSST19}. For $g>g_c$, the theory confines, and after including the spectator Fermi surface, we eventually obtain a conventional Fermi liquid description of the overdoped cuprates. Our focus in the body of the paper will be for values $g \geq g_c$ where there is no Higgs condensate  

We can take the limit of strong quartic interactions $u_0 \rightarrow \infty$ without modifying universal properties, and this simplifies the analysis and allows comparison with previous large $N_h$ work without gauge fields \cite{CSY94,DPSS}. The coupling $u_1$ is important in distinguishing possible Higgs phases for $g<g_c$, but it will not play a significant role for $g \geq g_c$. The gauge coupling $g_a$ will play no direct role in the large $N_h$ computations in this paper.

The effective potential $V(H)$ is constrained by the SU(2) gauge symmetry, and a global O($N_h$) symmetry acting on the flavor indices $\ell,m$. In the models considered in Ref.~\onlinecite{SSST19}, the global symmetry is smaller, and arises from the action of the square lattice space group symmetry on the charge density wave (and other) order parameters. We have enhanced the space group symmetry to O($N_h$) for simplicity \cite{SSST19,SPSS20}. 

The second term in the action is the long-range interaction obtained by integrating out the large Fermi surface of electrons.
\beq
\mathcal{S}_f = - \frac{1}{2 N_h} \int d^d x d \tau d \tau' H_{a \ell} (x, \tau) H_{a m} (x, \tau) J_f (\tau - \tau') H_{b \ell} (x, \tau') H_{b m} (x, \tau') \label{Sf}
\eeq
The electrons couple to the gauge-invariant order parameters
\beq
Q_{\ell m} = H_{a \ell} H_{a m} - \frac{\delta_{\ell m}}{N_h} H_{an} H_{a n}, \label{defQ}
\eeq
and then integrating out the electrons leads to the index and spacetime structure in $\mathcal{S}_f$; this structure will be crucial to the appearance of bilocality. We are assuming here that $Q_{\ell m}$ correspond to order parameters at non-zero wavevectors, and in that case we expect \cite{hertz}
$J_f (\tau) \sim 1/\tau^2$
at large $\tau$, which is the Fourier transform of a $|\omega|$ frequency dependence. 
In the more complete model of Ref.~\onlinecite{SSST19}, some of the $Q_{\ell m}$ correspond to order parameters at zero momentum, in which case the corresponding $J_f$ will be different: it will have both space and time dependencies arising from the Fourier transform of $|\omega|/|{\bm k}|$. We will not consider this more complex case here.

Our computations with $\mathcal{S}_f$ require an ultra-violet (UV) cutoff, and we choose
\beq
J_f (\tau) = \frac{K}{\varkappa^2 + \tau^2} \label{Jftau}
\eeq
where $\varkappa$ is a short time cutoff. This
has a simple Fourier transform
\beq
\tilde{J}_f(\omega) = \frac{\pi K}{\varkappa} e^{-\varkappa |\omega|} \label{e3}
\eeq
We will use the form in (\ref{e3}) but with Matsubara frequencies
\beq
\tilde{J}_f(\omega_n) = \frac{\pi K}{\varkappa} e^{-\varkappa |\omega_n|}\,. \label{deftJ}
\eeq
So the $T>0$ form of $J_f (\tau)$ is 
\bea
J_f (\tau) &=& \frac{\pi K T \sinh(2 \varkappa \pi T)}{\varkappa  \left[ \cosh(2 \varkappa \pi T) - \cos (2 \pi T \tau) \right]}.
\eea

To obtain the large $N_f$ limit, we decouple $\mathcal{S}_f$ by introducing a bilocal field $C_{ab} (x, \tau, \tau')$, and the terms in $V(H)$ with local fields $B_0(x,\tau)$, $B_{1,ab}(x,\tau)$. In this manner, we obtain the partition function  
\begin{align}
\mathcal{Z} &= \int \mathcal{D} C_{ab} (x, \tau, \tau') \mathcal{D} B_0 (x, \tau) \mathcal{D} B_{1,ab} (x, \tau) \mathcal{D} H_{a \ell} e^{ - \mathcal{S}_f - \mathcal{S}_b} \nonumber \\ 
\mathcal{S}_f &= \int d^d x d \tau d \tau' \left[ \frac{N_h}{2} \frac{ \left[C_{ab} (x, \tau, \tau') \right]^2}{J_f (\tau - \tau')} - C_{ab} (x, \tau, \tau') H_{a \ell} (x, \tau) H_{b \ell} (x, \tau') \right], \nonumber \\
\mathcal{S}_b &= \frac{1}{2} \int d^d x d \tau \Biggl[ \left[\partial_\mu H_{a \ell} (x, \tau)\right]^2 
+ i B_{0} (x, \tau) \left( H_{a \ell} (x, \tau) H_{a \ell} (x, \tau) - \frac{3 N_h}{g} \right) 
+ \frac{ N_h \left[B_{0} (x, \tau) \right]^2}{4 u_0} \nonumber \\
&~~~~~~~~~~~~~~~~~~~ 
+ i B_{1,ab} (x, \tau) H_{a \ell} (x, \tau) H_{b \ell} (x, \tau) +\frac{ N_h \left[B_{1,ab} (x, \tau) \right]^2}{4 u_1}  \Biggr].
\label{Sf+Sb}
\end{align}
In the large $N_h$ limit, we integrate over the $H_{a \ell}$ and obtain an effective action for the $C_{ab}$, $B_0$, and $B_{1,ab}$ with a prefactor of $N_h$. Note that the bilocal field $C_{ab} (x, \tau_1, \tau')$ is included in this effective action. The large $N_h$ limit then involves the saddle point analysis of this action, which we present in the following sections. 

\section{Large $N_h$ limit}
\label{sec:saddle}

For the symmetric phase, at the large $N_h$ saddle point, we take the following gauge invariant ansatz
\begin{align}
\label {ansatzsym}
\notag C_{ab} (x, \tau, \tau') &= \delta_{ab} C(\tau - \tau'),\\
\notag i B_{0} (x, \tau) &=B_0,\\
i B_{1,ab} (x, \tau) &= \delta_{ab} B_1.
\end{align}
From (\ref{Sf+Sb}) we observe that
\beq
\frac{3 B_1}{2 u_1} = \frac{B_0}{2 u_0} + \frac{1}{g} = \frac{1}{N_h} \left\langle H_{a \ell}^2 \right \rangle
\eeq
We can therefore express $B_1$ in terms of $B_0$ everywhere, and only treat $B_0$ as an independent variable. Let us also introduce $\tilde C(\omega_n)$ as the Fourier transform of $C(\tau)$, and define the parameter
\beq
[\Delta(T)]^2 \equiv B_0 + B_1 - 2 \tilde{C}(0)\,,
\eeq
where we explicitly identify the $T$ dependence to distinguish it from $\Delta \equiv \Delta(T=0)$.
We will see below that $[\Delta(T)]^{-1}$ is best understood as a spatial correlation length $\xi_x$, and not a temporal correlation length $\xi_\tau$; hence we do not call it a `gap'.
Then, in the limit $u_0 \rightarrow \infty$ the free energy density $F$ is a functional only of $\Delta(T)$ and $C(\tau)$ given by (after dropping an additive constant)
\bea
\frac{F[\Delta(T), C(\tau)]}{3N_h} &=& \frac{1}{2}\int_0^\beta d\tau \frac{\left[C(\tau)\right]^2}{J_f(\tau)} + \frac{T}{2} \sum_{\omega_n} \int^\Lambda \frac{d^2 k}{4 \pi^2} \ln
\bigl[k^2 + \omega_n^2 + [\Delta(T)]^2 - 2\tilde C(\omega_n) + 2\tilde C (0)\bigr] \nonumber \\
&~&~~~~~~~~~~~~~~~~~~ - \frac{[\Delta(T)]^2 + 2 \tilde{C}(0)}{2g} \,. \label{FDC}
\eea
Here $\beta = 1/T$, and $\Lambda$ is large momentum cutoff which we impose by a Pauli-Villars subtraction (see Section~\ref{sec:numerics}).
Our task in this section is to solve the saddle-point equations of $F$, and then determine $F$ as a function of $T$ and $g$.

The saddle point equations of (\ref{FDC}) are
\bea
C(\tau) &=& J_f (\tau) \int^\Lambda \frac{d^2 k}{4 \pi^2} G(k,\tau) \label{s1} \\
\frac{1}{g} &=& \int^\Lambda \frac{d^2 k}{4 \pi^2} G(k,0) \label{s2}
\eea
where the Higgs field Green's function is
\beq
\tilde G(k, \omega_n) = \frac{1}{k^2 + \omega_n^2 + [\Delta(T)]^2 - 2 \tilde{C} (\omega_n) + 2 \tilde{C}(0)}
\label{Gk}
\eeq
and $G(k, \tau)$ is its Fourier transform in frequency/time. We have to solve (\ref{s1}) and (\ref{s2}) for $\Delta (T)$ and $C(\tau)$ as a function of $T$ and $g$. In practice, it is easier to pick a value of $\Delta (T)$, solve (\ref{s1}) for $C(\tau)$, and then determine the value of $g$ as a dependent variable from (\ref{s2}). In particular, the critical value $g_c$ is determined by following this procedure for $\Delta(T=0) = 0$.

\subsection{Critical point}
\label{sec:Crit}

First, let us examine the nature of the critical point at $g=g_c$ at $T=0$.
Let us assume the power-law behavior
\beq
C(\tau) = \frac{\kappa_0}{|\tau|^{\alpha}} \quad \mbox{as $|\tau| \rightarrow \infty$}, \label{Ccrit}
\eeq
for some exponent $\alpha$ and prefactor $\kappa_0$.
Then
\beq
\tilde{C}(\omega) - \tilde{C}(0) = 2 \kappa_0 |\omega|^{\alpha-1} \Gamma( 1- \alpha) \sin (\pi \alpha/2) \quad \mbox{as $|\omega| \rightarrow 0$}.
\eeq
We can drop the $\omega_n^2$ term in (\ref{Gk}) if $\alpha < 3$. So we evaluate
\bea
\label{GtauAsymp}
\int \frac{d^d k d \omega}{(2 \pi)^{d+1}} \tilde G(k, \omega) e^{- i \omega \tau} &\approx& \int \frac{d^d k d \omega}{(2 \pi)^{d+1}}
\frac{e^{-i \omega \tau}}{k^2 - 4 \kappa_0 |\omega|^{\alpha-1} \Gamma( 1- \alpha) \sin (\pi \alpha/2)} \\
&=& \frac{\Gamma(1-d/2)}{(4 \pi)^{d/2}}\int \frac{d \omega}{2 \pi} e^{-i \omega \tau} \left[ - 4 \kappa_0 |\omega|^{\alpha-1} \Gamma( 1- \alpha) \sin (\pi \alpha/2) \right]^{(d-2)/2} \nn
&=& - \frac{\Gamma(1-d/2)}{(4 \pi)^{d/2}} \left[ - 4 \kappa_0 \Gamma( 1- \alpha) \sin (\pi \alpha/2) \right]^{(d-2)/2} \left[
\frac{\Gamma (1 + \delta) \sin (\pi \delta/2)}{\pi |\tau|^{1 + \delta}} \right] \nonumber
\eea
where 
\beq
\delta \equiv \frac{(\alpha - 1)(d-2)}{2}
\eeq
From saddle point equation (\ref{s1}), we now see that $\delta + 3 = \alpha$ or
\bea
\alpha = \frac{8-d}{4-d} = 3 - (2-d) + \ldots \quad \mbox{as $d \rightarrow 2$}\,,
\eea
and
\bea
\kappa_0 &=& \left( - K \frac{\Gamma(1-d/2)}{(4 \pi)^{d/2}} \left[ - 4 \Gamma( 1- \alpha) \sin (\pi \alpha/2) \right]^{(d-2)/2} \left[
\frac{\Gamma (1 + \delta) \sin (\pi \delta/2)}{\pi } \right] \right)^{2/(4-d)} \nn
&=& \frac{K}{4 \pi} \quad \mbox{as $d \rightarrow 2$}.
\eea
So we have a well-behaved result in the limit $d \rightarrow 2$ of interest to us:
\beq
C (\tau) = \frac{K}{4 \pi |\tau|^3} \quad \mbox{for $d=2$}\,.
\label{Ctaugc}
\eeq
Note that this result is linear in $K$, although we did {\it not\/} make an expansion in $K$ above; our analysis was only a low frequency asymptotic analysis. It can now be verified that computing the term linear in $K$ from (\ref{s1}) by dropping the $\tilde{C}$ contribution to $G$ on the right-hand-side also yields (\ref{Ctaugc}).

For the frequency dependence, the above results imply
\beq
\tilde{C}(0) - \tilde{C} (\omega) =\frac{K \omega^2}{4 \pi (2-d)} \quad \mbox{as $d \rightarrow 2$}.
\eeq
By the usual interpretation of dimensional regularization, we conclude
\beq
\label{C0log}
\tilde{C}(0) - \tilde{C}(\omega) = \frac{K \omega^2 \ln (\Lambda/|\omega|)}{4 \pi} \quad \mbox{for $d =2$}.
\eeq

\subsubsection{Subleading terms}
\label{sec:subleading}

It is useful to examine the structure of the subleading corrections to (\ref{C0log}) at low frequency, along with their dependence on $K$. Inserting (\ref{C0log}) into the right-hand-side of (\ref{s1}), transforming to frequency space, and performing the momentum integral, we obtain
\beq
\tilde{C} (\omega) =  \frac{K}{4} \int \frac{d \Omega}{2 \pi} |\omega-\Omega| \ln \left[ \frac{K \Omega^2 \ln (\Lambda/|\Omega|)}{2 \pi \Lambda^2}\right]
\eeq
Taking a derivative, we have for $\omega > 0$
\bea
\frac{d \tilde{C} (\omega)}{d \omega} &=&  \frac{K}{4} \int \frac{d \Omega}{2 \pi} \mbox{sgn}(\omega - \Omega) \ln \left[ \frac{K \Omega^2 \ln (\Lambda/|\Omega|)}{2 \pi \Lambda^2}\right] \nonumber \\
&\approx & \frac{K}{4 \pi} \omega \ln \left[ \frac{K \omega^2 \ln (\Lambda/|\omega|)}{2 \pi \Lambda^2}\right]
\eea
This agrees with (\ref{C0log}), and yields a subleading correction which is suppressed by a factor of
$\sim \ln[K \ln (\Lambda/|\omega|)]/\ln(\Lambda/|\omega|)$. We expect similar $\ln\ln/\ln$ corrections to all other aspects of the critical behavior to be discussed below.

\subsection{Fermi liquid regime}
\label{sec:Fermiliquid}

Let us now increase $g$ above $g_c$ to gap the Higgs field in the Fermi liquid phase at $T=0$.
We generalize the ansatz for $C(\tau)$ in (\ref{Ccrit}) to
\beq
\label{DisAsym}
C(\tau) = \frac{\kappa_1 e^{-|\tau|/\xi_\tau}}{|\tau|^{3}} \quad \mbox{as $|\tau| \rightarrow \infty$}.
\eeq
Then by Fourier transform,
\bea
\tilde{C}(0)-\tilde{C}(\omega)&=& \frac{\kappa_1}{\sqrt{2\pi}} \ln\left(\xi_\tau e^{-\gamma}\right) \omega^2 + O(\omega^4)\ \ \ \  \mbox{as $|\omega| \rightarrow 0$}.
\eea

We now confirm via the saddle point equations that the ansatz (\ref{DisAsym}) is self-consistent. 
For convenience, define $\tilde{\kappa}_1\equiv \kappa_1 \ln\left(\xi_\tau e^{-\gamma}\right)/\sqrt{2\pi}$. We approximate the Greens function in the small $\omega$ limit, and again evaluate using dimensional regularization (we define $\Delta \equiv \Delta(T=0)$),
\bea
\label{a:QDansatz}
\notag \int \frac{d^d k d \omega}{(2 \pi)^{d+1}} G(k, \omega) e^{- i \omega \tau} &\approx& \int \frac{d^d k d \omega}{(2 \pi)^{d+1}}
\frac{e^{-i \omega \tau}}{k^2 +\omega^2 + \Delta^2 + 2 \tilde{\kappa}_1 \omega^2} \\
&=& \frac{\Gamma(1-d/2)}{(4 \pi)^{d/2}}\int \frac{d \omega}{2 \pi} e^{-i \omega \tau} \left[\omega^2(1+2\tilde{\kappa}_1) + \Delta^2 \right]^{(d-2)/2} \nn
&=&\left[\frac{2 \left(2\Delta^2\right)^{(d/2-1) }  \left(\frac{1+2
   \tilde{\kappa}_1}{\Delta^2}\right)^{\frac{1}{4} (2 (d/2-1) -1)} K_{(d/2-1) +\frac{1}{2}}\left(\left| \tau
   \right|\sqrt{\frac{\Delta^2}{1+2 \tilde{\kappa}_1}}\right)}{(4 \pi)^{d/2}\left| \tau \right| ^{(d/2-1) +\frac{1}{2}}}  \right] \nonumber \\
 &=& \frac{1}{2\sqrt{2\pi}}\frac{\sqrt{2 \pi } e^{-\left| \tau \right|\sqrt{\frac{\Delta^2}{1+2 \tilde{\kappa}_1}}}}{\left| \tau
   \right| }\ \ \ \  \mbox{at $d=2$}.
\eea
Here, $K_n$ is a Bessel function.
Hence, asymptotically ($\tau\to\infty$) we have that
\bea
C(\tau)&=& J_f(\tau)G(\bm x=\bm 0,\tau)= \frac{K}{2\sqrt{2\pi}}\frac{e^{-\left| \tau \right|\sqrt{\frac{\Delta^2}{1+2 \tilde{\kappa}_1}}}}{\left| \tau
   \right|^3 }, \label{Ctau10}
   \eea
 which is consistent with (\ref{DisAsym}) once we identify $\kappa_1 = K/\sqrt{8\pi}$ and
   \bea
    \xi_\tau &=& \sqrt{1+K \ln\left(\xi_\tau e^{-\gamma}\right)/(2\pi)} \Delta^{-1} \nonumber \\
    & \approx &  \frac{1}{\Delta} \left[ \frac{K}{2 \pi} \ln (\Lambda/\Delta) \right]^{1/2} \label{xitau1}
   \eea
to leading logs.

We now determine the dependence of $\Delta$ on $(g-g_c)$. We obtain this by writing the difference of (\ref{s2}) at $g=g_c$ and $g>g_c$ as
\beq
\int\frac{d \omega}{2 \pi} \int^\Lambda \frac{d^2 k}{4 \pi^2} \left[
\frac{1}{k^2 + \omega^2 K \ln(\bullet)/(2\pi)} - \frac{1}{k^2 + \Delta^2 + \omega^2 K \ln(\bullet)/(2\pi)} \right] = \frac{1}{g_c} - \frac{1}{g} \,. \label{Delta1}
\eeq
Here $\ln(\bullet)$ refers to a logarithm of various possible frequency scales. In the leading logarithm approximation discussed in Section~\ref{sec:subleading}, we can just replace the logarithm by a constant with $\bullet = \Lambda/$(largest of external frequency scales); it can be verified that all of the results obtained so far in this section can also be obtained in this manner. Then (\ref{Delta1}) yields
\beq
\frac{\Delta}{4 \pi} \left[ \frac{K}{2 \pi} \ln (\Lambda/\Delta) \right]^{-1/2} = \frac{1}{g_c} - \frac{1}{g} \label{Deltag}
\eeq
or $\Delta \sim (g - g_c) \ln^{1/2}(1/(g-g_c))$. 

From (\ref{xitau1}) and (\ref{Deltag}), we see that $\xi_\tau^{-1} \sim (g-g_c)$, without a logarithmic correction. The absence of logarithmic corrections in $\xi_\tau$ will be crucial to our results. Also, from the structure of the Green's function, we see that the spatial correlation length, $\xi_x \sim \Delta^{-1}$. So there is a logarithmic singularity in the spatial correlation length, $\xi_x$,
but {\it not\/} in the temporal correlation length $\xi_\tau$.

\subsection{Non-zero temperatures}
\label{sec:nonzeroT}

The solution at $T>0$ is characterized by the parameter $\Delta (T)$. We can determine $\Delta(T)$ in terms of $\Delta \equiv \Delta(T=0)$, the parameter at the {\it same} value $g$ at $T=0$; the method leading to (\ref{Delta1}) now yields
\beq
 \int^\Lambda \frac{d^2 k}{4 \pi^2} \left[\int\frac{d \omega}{2 \pi}
\frac{1}{k^2  + \Delta^2 + \omega^2 K \ln(\bullet)/(2\pi)} - T \sum_{\omega_n} \frac{1}{k^2 + [\Delta(T)]^2 + \omega_n^2 K \ln(\bullet)/(2\pi)} \right] = 0 \,. \label{Delta2}
\eeq
This yields an equation for $\Delta (T)$ which is the same as that in Ref.~\onlinecite{CSY94} apart from the $\ln(\bullet)$ factors:
\beq
\frac{\Delta(T)}{T} \left[ \frac{K}{2 \pi} \ln (\bullet) \right]^{-1/2} = \Psi_\Delta \left(
\frac{\Delta}{T} \left[ \frac{K}{2 \pi} \ln (\bullet) \right]^{-1/2}\right)\,, \label{Deltascale}
\eeq
where the scaling function $\Psi_\Delta$ is the same as Ref.~\onlinecite{CSY94}
\beq
\Psi_\Delta (y) = 2 \, \mbox{arcsinh} \left(\frac{e^{y/2}}{2} \right)\,. \label{PsiDelta}
\eeq
In particular, at the critical point $g=g_c$, we have $\Delta=0$ and
\beq
\Delta(T) = \Theta T \left[ \frac{K}{2 \pi} \ln \left(\frac{\Lambda}{T} \right) \right]^{1/2} \label{deltaTlog}
\eeq
where $\Theta = 2 \ln((\sqrt{5}+1)/2)$. 

By combining (\ref{Deltascale}) with (\ref{xitau1}), we see that there is no logarithmic prefactor in the time-correlation length, as we observed above
\beq
\xi_\tau^{-1} (T) = T \Psi_\Delta \left(
\frac{\Delta}{T} \left[ \frac{K}{2 \pi} \ln (\bullet) \right]^{-1/2}\right)\,, \label{xitau2}
\eeq
so that $\xi_\tau (T) = 1/(\Theta T)$ at the critical point $g=g_c$, just as in Ref.~\onlinecite{CSY94}.
For the function $C (\tau)$, the leading-log corrections can be absorbed into $\xi_\tau$, and we expect from (\ref{Ctau10}) that
\beq
C (\tau) = \frac{1}{|\tau|^3} \Psi_C \left( \tau/\xi_\tau \right)\,, \label{Ctau11}
\eeq
with a scaling function $\Psi_C$.

\subsection{Free energy}
\label{sec:freeen}

As we will describe below, evaluating the free energy in (\ref{FDC}) leads to subtle questions on the nature of the low $T$ limit.
It turns out to be essential to have full analytical control in order to separate terms with different physical origins. 
We already observed below (\ref{Ctaugc}) that perturbation theory in $K$ was sufficient in determining the asymptotic form of $C(\tau)$. And we will see in Section~\ref{sec:rg} the critical coupling $K \sim 1/N_h$, which also justifies working at small $K$. We therefore divide the free energy as
\beq
F = F_H + F_K \label{Fdef}
\eeq
where $F_H$ is the large-$N$ contribution of the Higgs field, and $F_K$ contains contributions from the large Fermi surface which are first order in $K$. The analysis below amounts to an expansion in the free energy to linear order in $K$ about the critical point $g=g_c$. However the value of $g_c$ itself depends upon $K$, and this effect has to be treated more carefully.

\subsubsection{Evaluation of $F_H$}
\label{sec:FH}

At zeroth order in $K$, the Higgs field contribution in (\ref{FDC}) is the same as the free energy computed in Ref.~\onlinecite{CSY94}.
\beq
\frac{F_H ( 1/g,T) }{3N_h} =  \frac{T}{2} \sum_{\omega_n} \int^\Lambda \frac{d^2 k}{4 \pi^2} \ln
\bigl[k^2 + \omega_n^2 + [\Delta(T)]^2\bigr]  - \frac{[\Delta(T)]^2}{2g} \,. \label{CSY1}
\eeq
In this expression (and in the remainder of Section~\ref{sec:freeen}), it is implied that $\Delta (T)$ is to be evaluated at the saddle point of (\ref{CSY1}) with respect to $\Delta (T)$. We write the saddle point equation (\ref{s2}), in a manner analogous to (\ref{Delta2}), as
\beq
\int^\Lambda \frac{d^2 k}{4 \pi^2} \left( T \sum_{\omega_n} \frac{1}{k^2 + \omega_n^2 + [\Delta (T)]^2} - 
\int \frac{d \omega}{2 \pi} \frac{1}{k^2 + \omega^2} \right) = \frac{1}{g} - \frac{1}{g_c^{0}} \label{CSY11}
\eeq
where $g_c^{0} = 4 \pi/\Lambda$ is critical coupling at which $\Delta (T=0) = 0$.

However, before inserting the evaluation of (\ref{CSY1}) into (\ref{Fdef}), we have to apply a renormalization procedure which is entirely analogous to converting perturbative field-theoretic expansions of critical phenomena from `bare' mass propagators to `renormalized' mass propagators, in which some perturbative terms are included to all orders \cite{amit1984field}. 
Here, this is essential for ensuring that our perturbative results in powers of $K$ hold not only in the perturbative regime where $K \Lambda \ll T$, but also in the limit $T \rightarrow 0$ at the critical point. In the present context, the `mass' is the distance of coupling $1/g$ from the critical point $1/g_c$. At first order in $K$ there is a correction to the value of $1/g_c$ which is computed in Appendix~\ref{app:gc} to be
\beq
\frac{1}{g_c} = \frac{1}{g_{c}^{0}} + \frac{1}{g_c^{1}} \quad; \quad \frac{1}{g_c^{1}} = - \frac{K \Lambda}{8 \pi^2} \left( 2 \ln 2 - \frac{1}{2} \right)\,, \quad \mbox{as $\varkappa \rightarrow 0$.} \label{CSY12}
\eeq
So we introduce a `renormalized' coupling $1/g_R$ related to the `bare' coupling $1/g$ via
\beq
\frac{1}{g_R} - \frac{1}{g_c^{0}} = \frac{1}{g} - \frac{1}{g_c} \quad \Rightarrow \quad \frac{1}{g} = \frac{1}{g_R} + \frac{1}{g_c^1} \,. \label{CSY13}
\eeq
Now when we evaluate the zeroth order expressions (\ref{CSY1}) and (\ref{CSY11}) at the renormalized coupling $1/g_R$ we obtain at $T=0$
\beq
\int^\Lambda \frac{d^2 k}{4 \pi^2} \int \frac{d \omega}{2 \pi} \left[ \frac{1}{k^2 + \omega^2 + \Delta^2} - 
 \frac{1}{k^2 + \omega^2}  \right] = \frac{1}{g_R} - \frac{1}{g_c^{0}}\,. \label{CSY11a}
\eeq
Now we see from (\ref{CSY13}) that $\Delta=0$ precisely when the renormalized `mass' $1/g_R - 1/g_c^0$ vanishes at the true critical point where $g=g_c$. After the substitution in (\ref{CSY13}), we treat $1/g_R$ as independent of $K$, and expand everything in powers of the perturbative coupling $K$, just as is done in renormalized mass expansions in field theory. So we need to evaluate $F_H (1/g,T) = F_H(1/g_R + 1/g_c^1, T)$ to linear order in $K$. Here,
we are aided by the fact that $F_H$ is a saddle point with respect to variations in $\Delta (T)$, so we need not account for the shift in $\Delta (T)$ to linear order. Indeed, we need only consider the variation arising from the only explicit linear dependence of (\ref{CSY1}) on $1/g$. 
So we have from (\ref{CSY1}) and (\ref{CSY13})
\beq
\label{FHshift}
F_H (1/g, T) = F_H (1/g_R, T)- \frac{[\Delta(T)]^2}{2g_c^1}
\eeq
Finally, we collect together the results of Section~\ref{sec:FH}, perform the frequency summations in (\ref{CSY1}) and (\ref{CSY11}) to obtain the final expression for $F_H$
\bea
\frac{F_H}{3N_h} &=& \frac{\Lambda^3}{12 \pi} + \int^\Lambda \frac{d^2 k}{4 \pi^2} \Biggl[
\frac{1}{2} \sqrt{k^2 + [\Delta(T)]^2} - \frac{k}{2} + T \ln \left( 1 - e^{- \sqrt{k^2 + [\Delta(T)]^2}/T} \right) \nonumber \\
&~&~- \frac{[\Delta(T)]^2}{4\sqrt{k^2 + [\Delta(T)]^2}}
\left\{ 1 + 2 n\left( \sqrt{k^2 + [\Delta(T)]^2} \right) \right\} \Biggr] - \frac{\left[ \Delta (T) \right]^2}{2 g_c^1}\,, \label{FHres}
\eea
where $n(a) = 1/(e^{a/T} - 1) $ is the Bose function, and the value of $\Delta(T)$ is related to $\Delta$ by
\beq
\int^\Lambda \frac{d^2 k}{4 \pi^2} \left[\frac{1}{2\sqrt{k^2 + [\Delta(T)]^2}}\left\{ 1 + 2 n\left( \sqrt{k^2 + [\Delta(T)]^2} \right) \right\} - \frac{1}{2 \sqrt{k^2 + \Delta^2}} \right] = 0\,. \label{CSY14}
\eeq
The integrals over $k$ are convergent as $\Lambda \rightarrow \infty$ in both (\ref{FHres}) and (\ref{CSY14}). Consequently, the corresponding contribution to the free energy scales as $T^3 \Phi_H (\Delta/T)$, where the scaling function $\Phi_H$ was given in Ref.~\onlinecite{CSY94}. The renormalized coupling $g_R$ appears only in determining the value of $\Delta$ in (\ref{CSY11a}), and we will express all remaining results in this section in terms of $\Delta$.
In $F_K$ we can simply replace $1/g$ by $1/g_R$, because those terms are already first order in $K$.

\subsubsection{Evaluation of $F_K$}

We turn next to the term $F_K$ in $F$, which contains all terms which are linear in $K$ corrections to $F_H$ in (\ref{CSY1}).
We can obtain these terms simply by evaluating the expectation value of $\mathcal{S}_f$ in (\ref{Sf}) in the large $N_h$ limit. 
So we obtain
\beq
\frac{F_K}{3 N_h} = - \frac{1}{2} \int_0^{\beta} d \tau J_f (\tau) G^2(x=0, \tau)
\eeq
or in frequency space
\beq
\frac{F_K}{3 N_h}= -\frac{1}{2} T^2 \sum_{\omega_n, \epsilon_n} \tilde J_f (\omega_n) \left[\int^\Lambda \frac{d^2k}{(2\pi)^2} \tilde G(k,\epsilon_n) \right]\left[\int^\Lambda \frac{d^2p}{(2\pi)^2} \tilde G(p,\epsilon_n- \omega_n) \right] \label{e4}
\eeq
In (\ref{e4}) we use the Green's function in (\ref{Gk}) at zeroth order in $K$
\beq
\tilde G(k, \omega_n) = \frac{1}{k^2 + \omega_n^2 + [\Delta (T)]^2}\,. \label{e4a}
\eeq
To compute the free energy, we first evaluate the summation over $\epsilon_n$ in (\ref{e4}) using the identity
\beq
T \sum_{\epsilon_n} \frac{1}{(\epsilon_n^2 + a^2)((\epsilon_n - \omega_n)^2 + b^2)} = \frac{1}{2ab} \left[\frac{(b-a)(n(a) - n(b))}{\omega_n^2 + (a-b)^2} + 
\frac{(a+b)(1 + n(a) + n(b))}{\omega_n^2 + (a+b)^2} \right] \,. \label{e5}
\eeq
We also use (\ref{e3})
to evaluate by the contour integration method
\bea
T\sum_{\omega_n} \frac{\tilde J_f (\omega_n)}{\omega_n^2 + c^2}=  \frac{\pi K \cos(\varkappa c)}{2 \varkappa c} ( 1 + 2 n(c))
+ \frac{K}{\varkappa} \int_0^{\infty} d \Omega \, \mathcal{P} \left(\frac{1}{c^2 - \Omega^2} \right) \sin (\varkappa \Omega) (1 + 2 n (\Omega)) \label{e6}
\eea
Combining (\ref{e4},\ref{e5},\ref{e6}), and changing variables of integration from $k,p$ to $a = (k^2 + [\Delta(T)]^2)^{1/2}$, $b = (p^2 + [\Delta(T)]^2)^{1/2}$ we obtain
\bea
\frac{F_K}{3 N_h} &=& - \frac{1}{16\pi^2} \int_{\Delta (T)}^{\sqrt{\Lambda^2 + [\Delta(T)]^2}} da \int_{\Delta (T)}^{\sqrt{\Lambda^2 + [\Delta(T)]^2}} db \Biggl[
\frac{\pi K \cos(\varkappa (a-b))}{2 \varkappa} ( 1 + 2 n(a-b)) (n(b) - n(a)) \nonumber \\
&~&~~~~~~~+ \frac{\pi K \cos(\varkappa (a+b))}{2 \varkappa}  ( 1 + 2 n(a+b)) (1 + n(b) + n(a))  \label{e7} \\
&~&~~~~~~~+ \frac{K}{\varkappa} \int_0^{\infty} d \Omega \, \mathcal{P} \left(\frac{1}{(a-b)^2 - \Omega^2} \right) \sin (\varkappa \Omega) (1 + 2 n (\Omega)) (b-a)(n(a)-n(b)) \nonumber \\
&~&~~~~~~~+ \frac{K}{\varkappa} \int_0^{\infty} d \Omega \, \mathcal{P} \left(\frac{1}{(a+b)^2 - \Omega^2} \right) \sin (\varkappa \Omega)  (1 + 2 n (\Omega))(a+b) (1+ n(a)+n(b))  \Biggr]\,. \nonumber
\eea
Details of the evaluation of the integrals in (\ref{e7}) in the limit $\varkappa \rightarrow 0$ appear in Appendix~\ref{app:linearK}.
We can analytically evaluate the integrals while only dropping terms which scale as $T^3 \ln (\bullet)$ {\it and\/} are exponentially small in the regime $T \ll \Delta$. The omitted terms are argued to scale as $T^{d/z +1}$ in Section~\ref{sec:rg}, and they preserve hyperscaling; they will be
numerically evaluated in Section~\ref{sec:numerics}. In this approximation we find as $\varkappa \rightarrow 0$ (recall that $\Delta \equiv \Delta (T=0)$)
\bea
\frac{F_K}{3 N_h}
 &\approx&  - \frac{ K}{32 \pi \varkappa} \left(\sqrt{\Lambda^2+ \Delta^2} - \Delta \right)^2 - \frac{K T^2}{24} \left[ \Lambda \ln(2) -  \Delta \ln \left( \frac{\Lambda}{2\Delta} \right) - \Delta \right] \nonumber \\ &~&~~ - \frac{K \Lambda^3}{16 \pi^2} \left[ \ln (\varkappa \Lambda) + \frac{4}{3} \ln 2 - \frac{5}{6} \right] + \frac{K \Lambda^2 \Delta}{16 \pi^2}\left( 1 - \frac{\Delta}{2 \Lambda} \right) \left[\ln (\varkappa \Lambda ) - \frac{1}{2} \right] \nonumber \\
&~&~~ - \frac{K \Lambda [\Delta(T)]^2}{16\pi^2} \Biggl[ 2 \ln 2 - \frac{1}{2} \Biggr]\,. \label{FKres}
\eea

We now analyze the structure of the main result of this subsection in (\ref{FKres}). 
An important observation is that the term proportional to $\Lambda [\Delta (T)]^2$ cancels exactly with the corresponding term in (\ref{FHres}) which arose from the shift in $g_c$ to linear order in $K$, as shown in (\ref{CSY12}) and computed in (\ref{gc1}) in Appendix~\ref{app:gc}. This term is proportional to $\Lambda$, and so could have led to hyperscaling violation. It is remarkable that all hyperscaling violating terms exactly cancel: the mechanics of this cancellation is described in Appendix~\ref{app:gc}.

From the remaining temperature dependent terms in (\ref{FKres}), we therefore obtain a simple expression for the specific heat
\beq
\frac{C_v}{3 N_h} = \frac{K T}{12} \left[ \Lambda \ln(2) -  \Delta \ln \left( \frac{\Lambda}{2\Delta} \right) - \Delta \right]\,. \label{CDelta}
\eeq
The term proportional to $\Lambda$ is independent of couplings, and so contributes to the background $\gamma_b$ term in (\ref{Cscale}): it can be viewed as a finite enhancement of the mass of the background fermions from the Higgs fluctuations. The remaining terms in (\ref{CDelta}) correspond to free energy scaling $T^3 \ln (\bullet)$, but are not exponentially small for $T \ll \Delta$: hence they were not dropped in Appendix~\ref{app:linearK}. These terms dominate the specific heat for $T \ll \Delta$, yielding a $\Delta$-dependent co-efficient for a linear-in-$T$ specific heat. As we will see in the renormalization group analysis in Section~\ref{sec:rg}, we expect $\Delta \ln (\Lambda/\Delta)$ to exponentiate to $\Delta^{d/z-1}$, and so (\ref{CDelta}) contributes to the hyperscaling preserving contribution term in (\ref{Cscale}). This is the dominant singular term contributing to $\lim_{T \rightarrow 0} C_v/T$.

\begin{figure}
\begin{center}
\includegraphics[width=3.2275in]{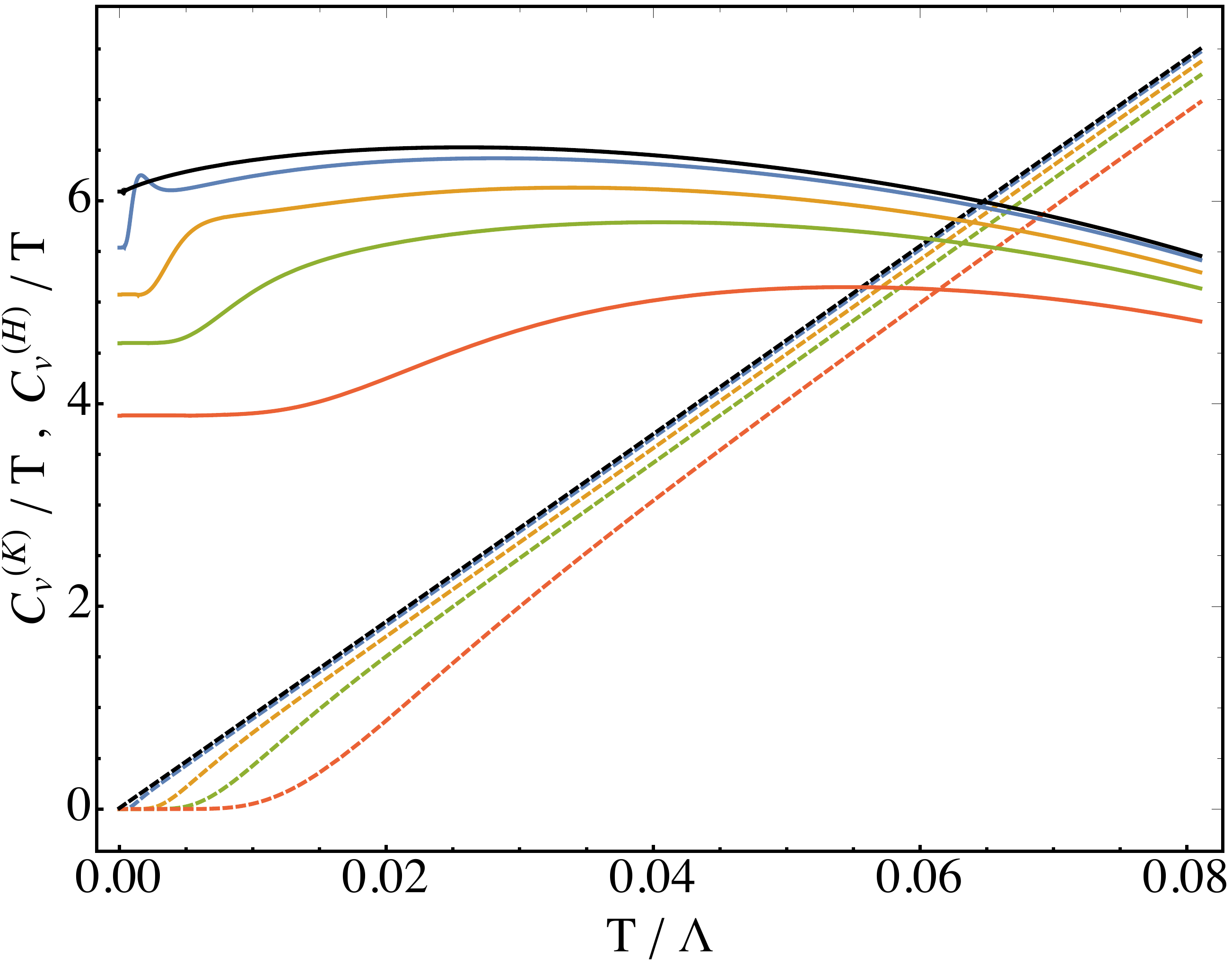} \ 
\includegraphics[width=3.3in]{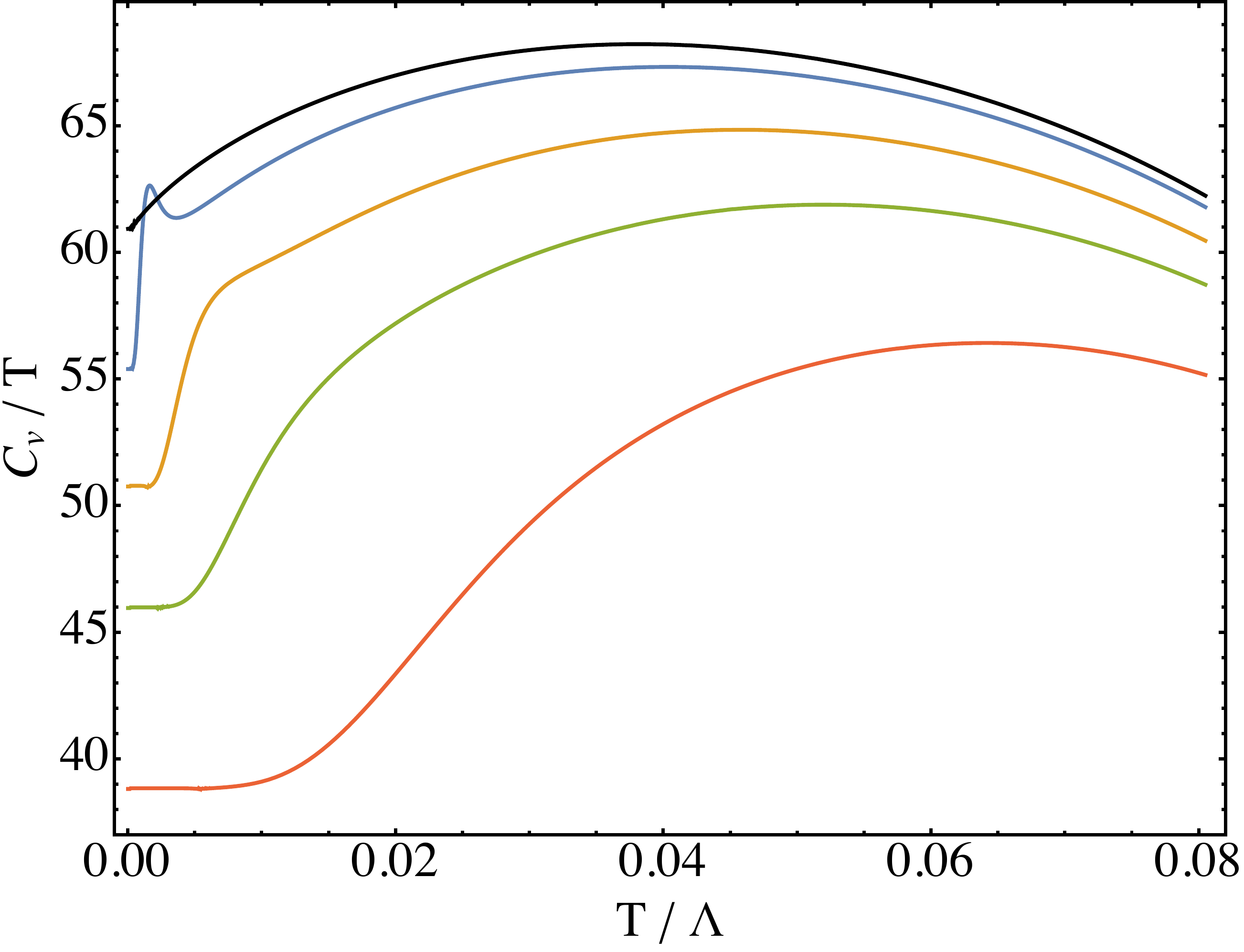}\\
\includegraphics[width=3.3in]{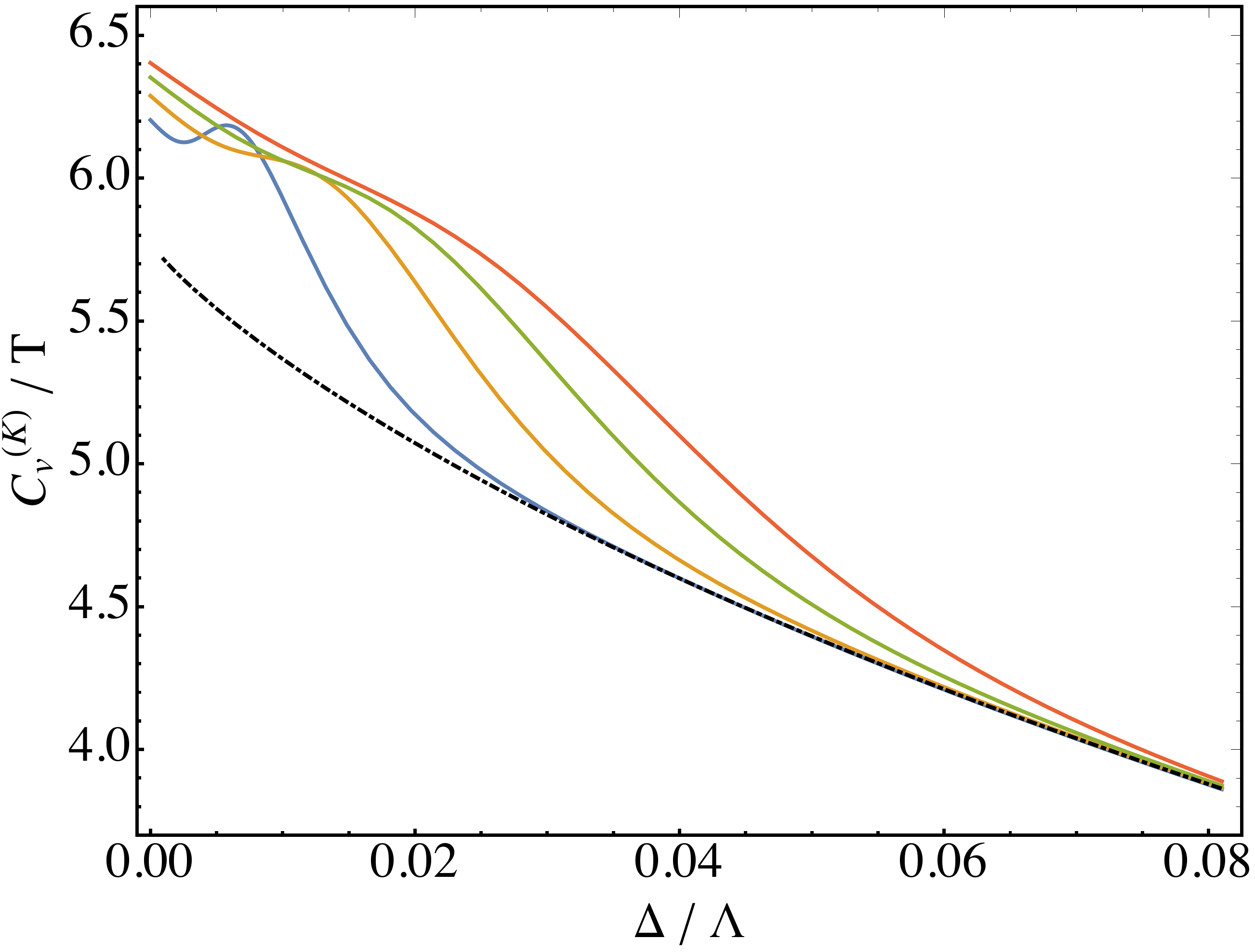} \ 
\includegraphics[width=3.25in]{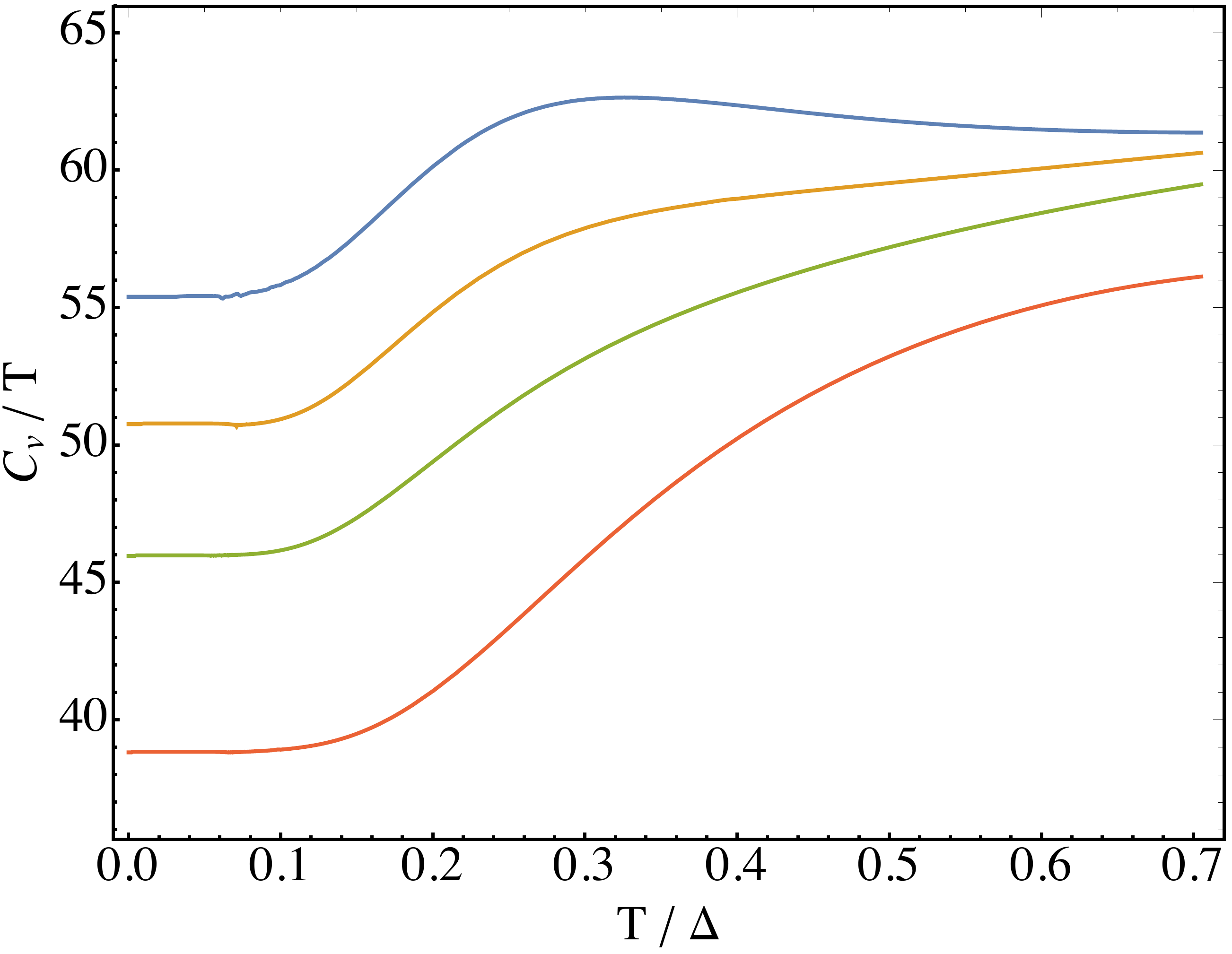}
\end{center}
\caption{Specific heat contributions $ C_v^{(H)}/T$ and $C_v^{(K)}/T$, evaluated from (\ref{CvHK}).  Everywhere $\varkappa=1/\Lambda$. (a) and (b) as a function of $T$ at fixed $\Delta$. Black, blue, orange, green and red correspond to $\Delta /\Lambda=\{0, 0.5, 2, 4, 8\}\times10^{-2}$. In (a) solid and dashed lines correspond to $C_v^{(K)}/T$ and $C_v^{(H)}/T$, and for presentation we take $K=1$. In (b) both contributions are summed $C_v/T = C_v^{(H)}/T+C_v^{(K)}/T$, and we take $K=10$. (c) $C_v^{(K)}/T$ as a function of $\Delta$ and fixed $T$: Blue, orange, green and red correspond to $T/ \Lambda=\{0.25, 0.5, 0.75, 1.0\}\times10^{-2}$. The black dot-dashed line corresponds to the asymptotic form obtained for the Fermi liquid regime in (\ref{CDelta}). (d) $C_v/T$ vs $T/\Delta(g)$, with $K=10$. Same color scheme as in (a) and (b).   
} \label{f:Cv}
\end{figure}

\begin{figure}
\begin{center}
\includegraphics[width=5.2in]{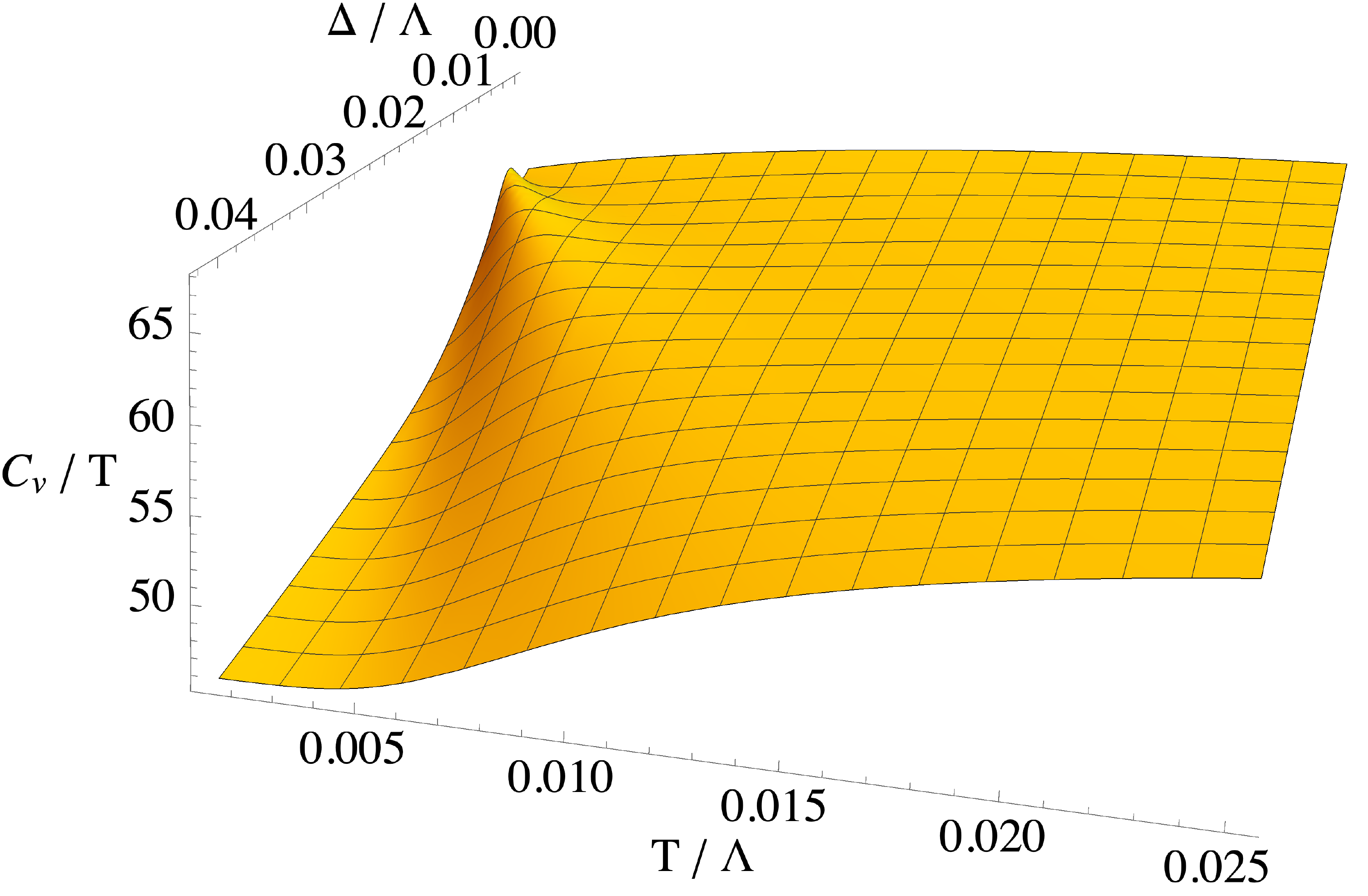}
\end{center}
\caption{Specific heat surface plot, $C_v/T = C_v^{(H)}/T+C_v^{(K)}/T$ versus $T/\Lambda$ and $\Delta/\Lambda$, with $K=10$.} \label{f:Cv3D}
\end{figure}

\subsection{Numerical solution}
\label{sec:numerics}
We now turn to a numerical evaluation of the free energy and specific heat in the $K$-expansion, and an evaluation of the Greens function and decoupling fields in the self-consistent theory, {\it i.e.\/} to all orders in $K$.

\subsubsection{First order in $K$}
Within the $K$-expansion, we focus on the features of the specific heat $C_v/T = C_v^{(H)}/T + C_v^{(K)}/T$, coming from the free energy contributions $F= F_H + F_K$, with $F_H$ in (\ref{FHres}) and $F_K$ in (\ref{e4}). However, we reshuffle such that all $K$ dependence is collected into $C_v^{(K)}/T$, which is achieved via
\begin{align}
\label{CvHK}
\notag C_v^{(H)}/T&\equiv - \frac{\partial^2}{\partial T^2}\left[F_H +  \frac{[\Delta(T)]^2}{2g_c^1}\right],\\
C_v^{(K)}/T&\equiv - \frac{\partial^2}{\partial T^2}\left[F_K -  \frac{[\Delta(T)]^2}{2g_c^1}\right].
\end{align}
For the evaluation, we take $\varkappa=1/\Lambda$, which requires we use the full expression for $g_c^1$ presented in (\ref{gc1}). The temperature dependence of $\Delta(T)$ is obtained from (\ref{CSY11}). 

Figure \ref{f:Cv}(a) and (b) looks at the relative and combined contributions of $C_v^{(H)}/T$ and $C_v^{(K)}/T$, as a function of $T$ at fixed $\Delta$. We see a non-monotonic dependence in $C_v/T$, coming from the contribution $C_v^{(K)}/T$, with a peak at a value $T \sim \Delta$ -- this is further manifest in Figure \ref{f:Cv}(d). Such a peak indicates the change of regime from Fermi liquid $\Delta\gg T$ to quantum critical $\Delta\ll T$, and as such could be a useful experimental \cite{Loram01,Michon18,Loram19} diagnostic of the critical point. In Figure \ref{f:Cv}(c) we plot $C_v^{(K)}/T$ versus $\Delta$ at fixed $T$, which demonstrates a significant conclusion of the present analysis; that upon tuning to the critical point $\Delta\to0$, $\lim_{T \rightarrow 0} C_v^{(K)}/T$ (and hence $\lim_{T \rightarrow 0} C_v/T$) is enhanced. Figure \ref{f:Cv3D} provides a surface plot of $C_v/T$ versus $T$ and $\Delta$, with fixed $K=10$.

\subsubsection{All orders in $K$}
We now present aspects of the theory obtained to all orders in $K$; namely the full bosonic mass gap $\Delta(1/g, K,T)$, Greens function $G(\tau)$, and the saddle point of the bilocal field, i.e. $C(\tau)$ and its Fourier transform $\tilde{C}(\omega_n)$. 

For the sake of a self-consistent numerical treatment, the Pauli-Villars procedure is ideal because it does not introduce any sharp cutoffs. Because we also need to regulate the free energy with the same procedure, we choose two subtractions:
\beq
\tilde G(k, \omega_n) = \frac{1}{k^2 + \Sigma (\omega_n)} + \frac{1}{k^2 + \Sigma (\omega_n) + 2 \Lambda^2} - \frac{2}{k^2 + \Sigma (\omega_n) + \Lambda^2}\,. \label{GPauli}
\eeq
where we have defined 
\beq
\Sigma(\omega_n) \equiv \omega_n^2 + \Delta^2 -2 \tilde C (\omega_n) + 2 \tilde C (0)\,.
\eeq
This ensures a $\sim (k^2 + \Sigma (\omega_n))^{-3}$ decay at large $k$ and $\omega_n$. 

\begin{figure}
\begin{center}
\includegraphics[width=3.4in]{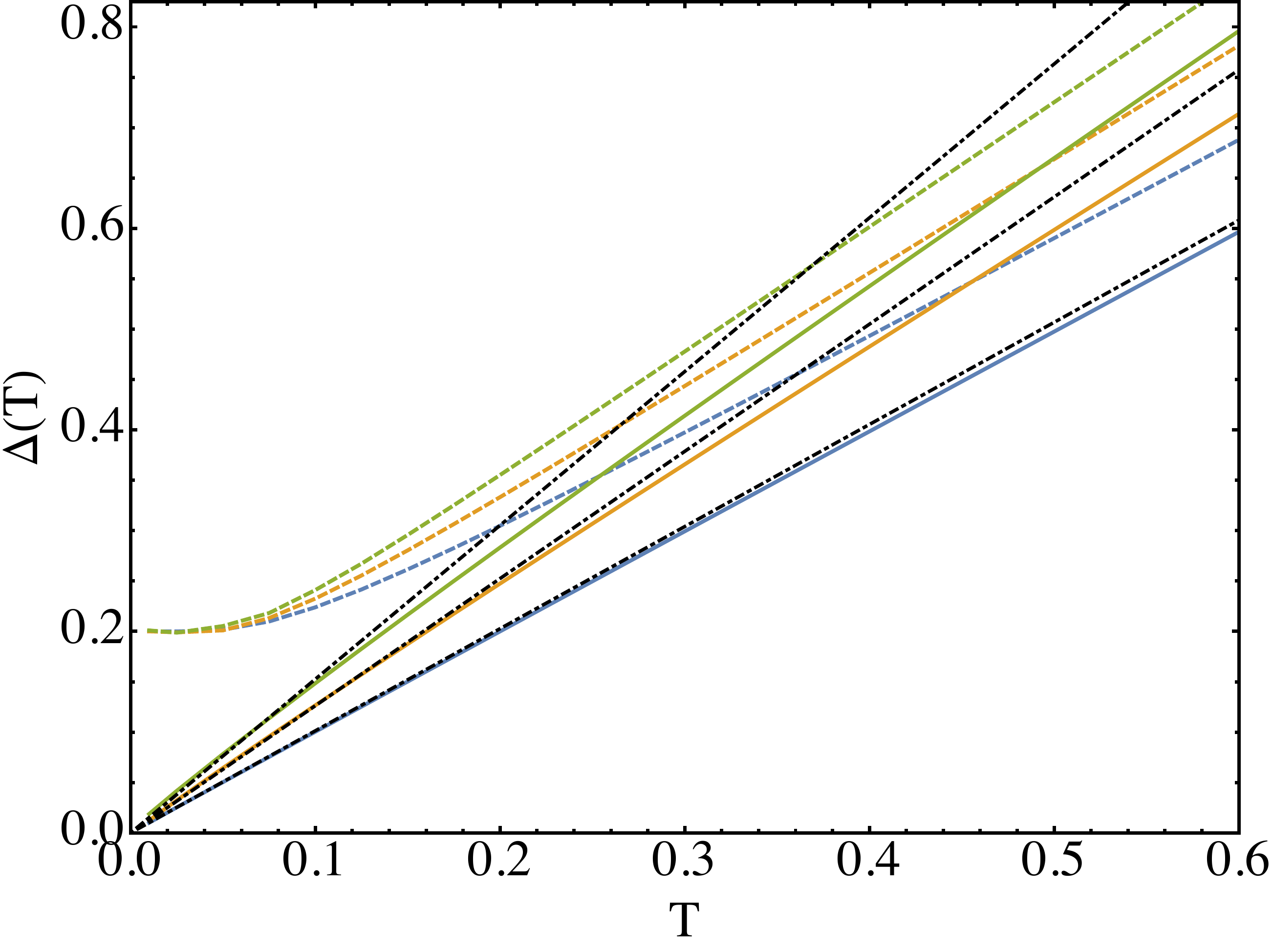}
\includegraphics[width=3.25in]{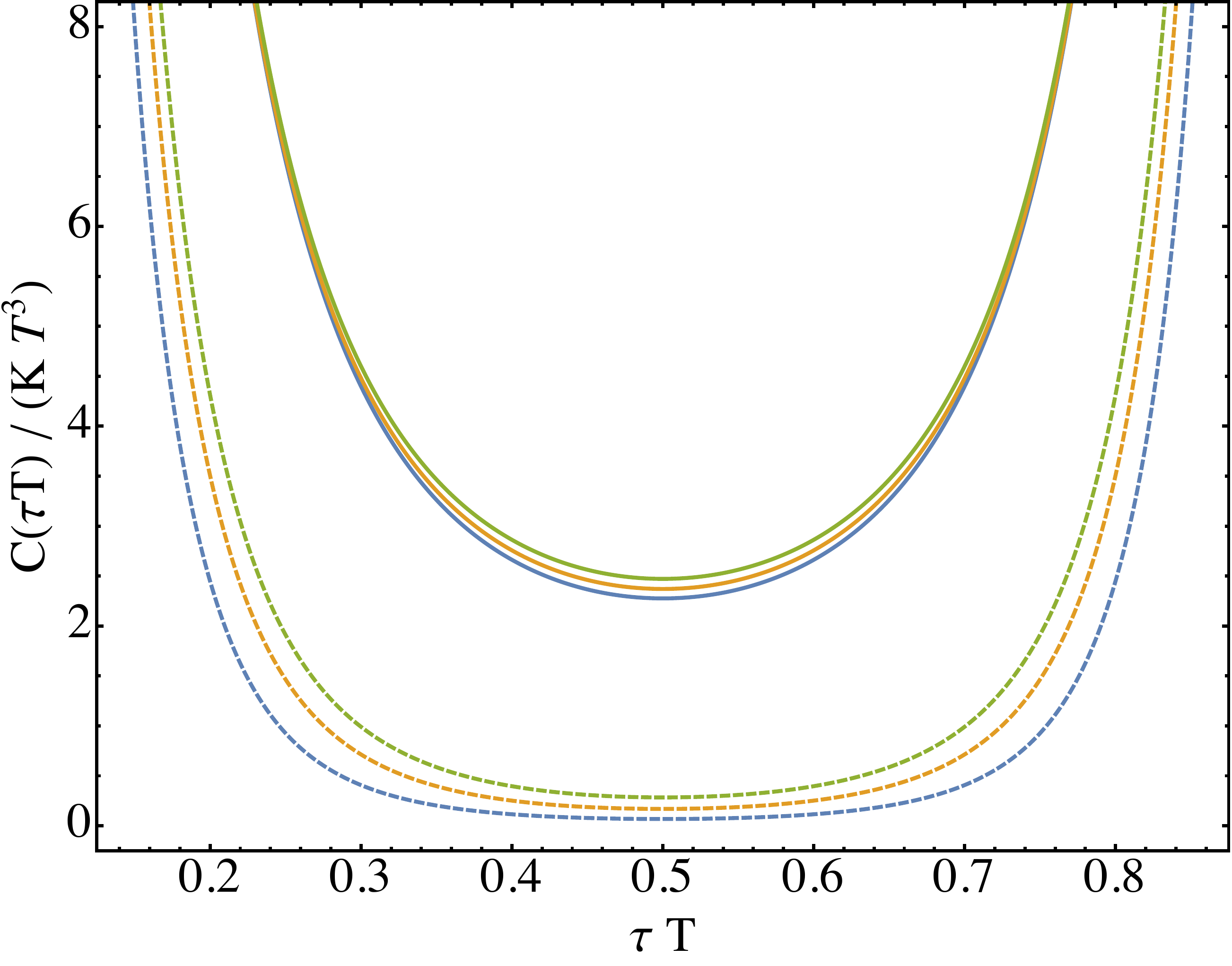}
\includegraphics[width=3.25in]{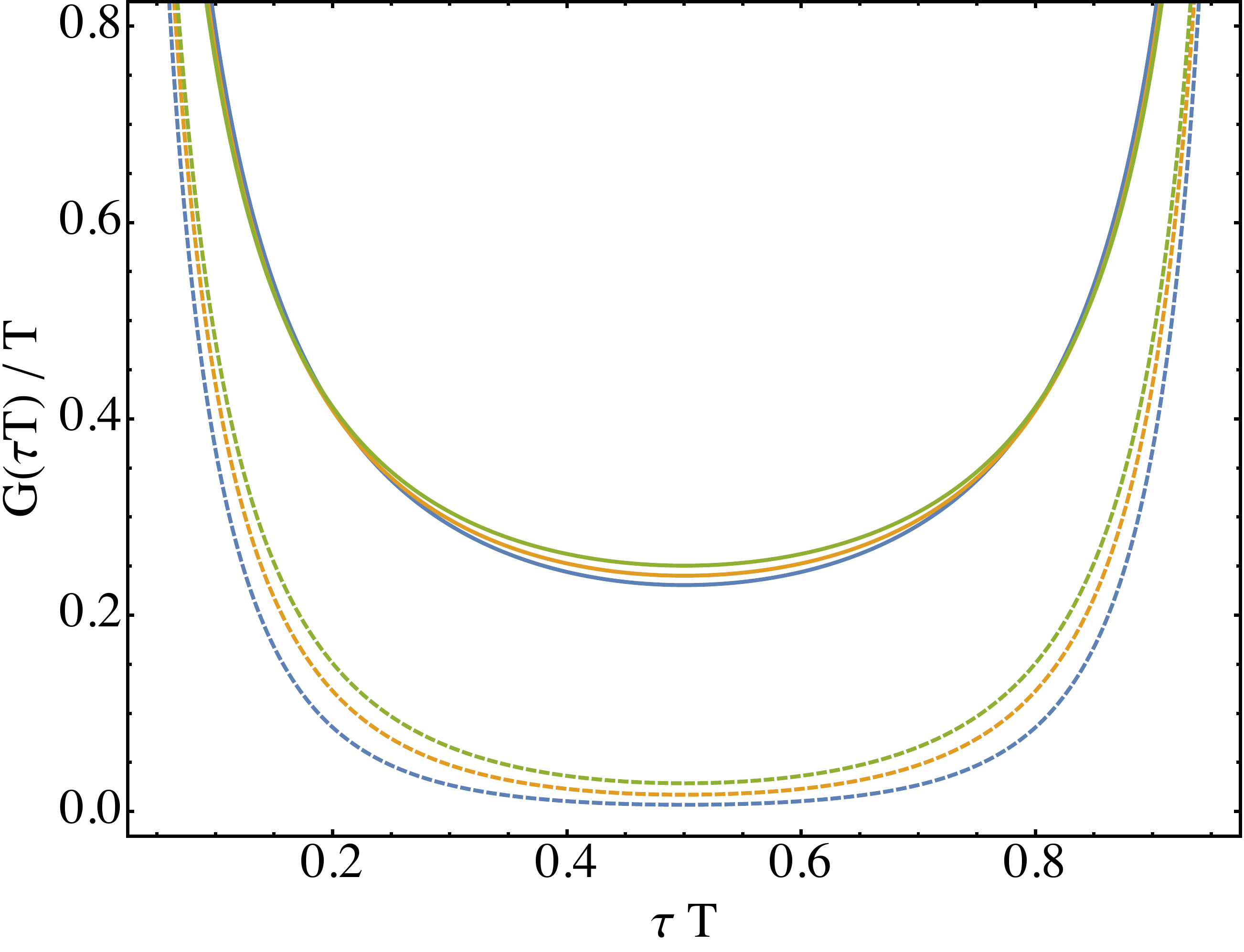} \hspace{0.1cm}
\includegraphics[width=3.35in]{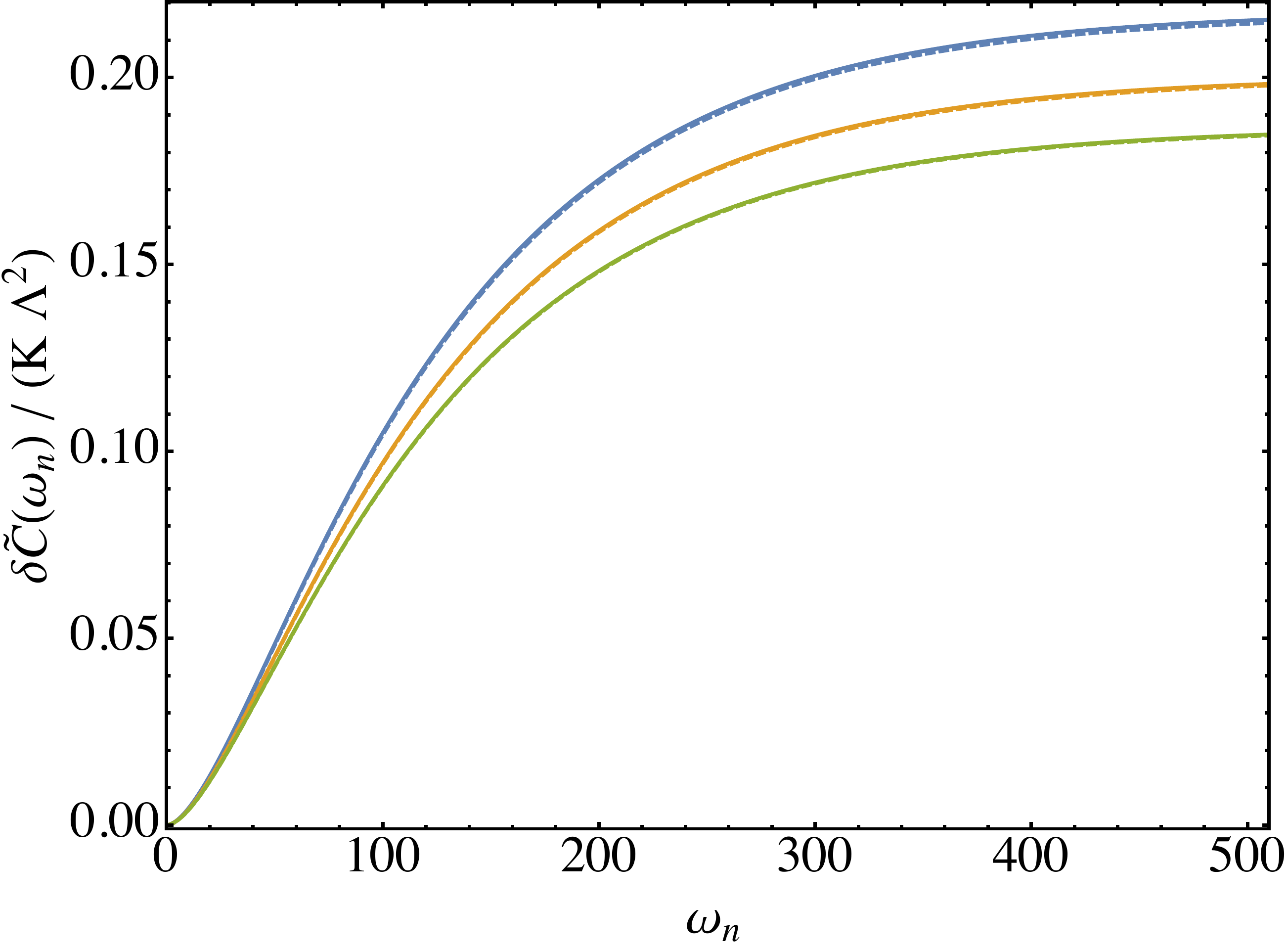}
\end{center}
\caption{{ Self-consistent saddle point solutions -- Blue, orange, green correspond to $ K=\{0.01,0.1,1\}$: Solid lines correspond to the quantum critical mass gap $\Delta(T)$ at $g=g_c$, while dashed corresponds to $g>g_c$ whereby the zero temperature gap $\Delta(T=0)/\Lambda=2\times10^{-3}$. (a) Inverse spatial correlation length, $\Delta(T,K) \sim \xi_x^{-1}$.  Dash-dotted black lines correspond to linear in $T$ fits at small $T$, and are merely a guide to the eye. The curves for $\Delta(T)$ at $g=g_c$ are consistent with log correction obtained analytically in (\ref{deltaTlog}), whereby at larger $K$ we expect larger log corrections. (b) $C(\tau T)/(K T^3)$ vs $\tau T$, (c) $G(\tau T)/T$ vs $\tau T$, and (d) 
$\delta \tilde{C}(\omega_n)/(K \Lambda^2 )$ vs $\omega_n$, where $\delta \tilde{C}(\omega_n) =2\tilde{C}(0) - 2\tilde{C}(\omega_n)$:  In all cases $T/\Lambda= 0.25\times 10^{-3}$.} \label{f:Saddle}}
\end{figure}

For the free energy, the corresponding regularization is
\bea 
&& \int^\Lambda \frac{d^2 k}{4 \pi^2} \ln
\bigl[k^2 + \Sigma (\omega_n) \bigr] = \nonumber \\
&&~~~\int \frac{d^2 k}{4 \pi^2} \Bigr[ \ln
\bigl[k^2 + \Sigma (\omega_n) \bigr] + \ln
\bigl[k^2 + \Sigma (\omega_n) + 2 \Lambda^2 \bigr]  - 2 \ln
\bigl[k^2 + \Sigma (\omega_n) + \Lambda^2 \bigr] \Bigr]
\eea
where the right-hand-side decays as $\sim (k^2 + \Sigma (\omega_n) )^{-2}$ at large $k$ and $\omega_n$, and the saddle point equations of the free energy yield (\ref{GPauli}). The $k$ integration is readily performed, and upon applying this regularization scheme to the free energy (\ref{FDC}), the corresponding
saddle point equations (\ref{s1}) and (\ref{s2}) become,
\bea
C(\tau) &=& J_f (\tau) \frac{T}{4 \pi} \sum_{\omega_n} e^{- i \omega_n \tau} \ln \left[
\frac{[\Sigma (\omega_n) + \Lambda^2]^2}{ \Sigma (\omega_n) [\Sigma (\omega_n) + 2 \Lambda^2]}\right]  \\
\frac{1}{g} &=& \frac{T}{4 \pi} \sum_{\omega_n} \ln \left[
\frac{[\Sigma (\omega_n) + \Lambda^2]^2}{ \Sigma (\omega_n) [\Sigma (\omega_n) + 2 \Lambda^2]}\right].
\eea
We provide the numerical solution of these saddle point equations in Figure \ref{f:Saddle}. There our focus is on the critical coupling $g=g_c$, as well as one value of $g>g_c$, chosen such that $\Delta(T=0)/\Lambda=2\times10^{-3}$. Having these two values of $g$ allows us to tease out the key qualitative features of the saddle point solutions. Figure \ref{f:Saddle}(a) shows the mass gap as a function of $T$. The $g=g_c$ results are consistent with log correction obtained analytically in (\ref{deltaTlog}). Figure \ref{f:Saddle}(b) and (c) test the logarithmically violated scaling in (\ref{Ctau11}), and show the non-linear influence of $K$ on $C(\tau)$ and $G(\tau)$ -- from the `large' time $\tau T \to 1/2$ asymptotic, we see that in the critical case $g=g_c$, the deviation from linearity in $K$ is likely only as weak as logarithmic in $K$. These figures also show the expected suppression of these functions for the case $g>g_c$, relative to the critical case $g=g_c$, which becomes especially pronounced at `large' times, $\tau T \to 1/2$. Finally, in Figure \ref{f:Saddle}(d) we show the frequency space behaviour of $\delta\tilde{C}(\omega_n) = 2\tilde{C}(0) - 2\tilde{C}(\omega_n)$.      

\section{Renormalization group analysis}
\label{sec:rg}

This section will explore the nature of the $1/N_h$ corrections to the $N_h = \infty$ theory presented in Section~\ref{sec:saddle}. A complete examination of such corrections requires determination of the fluctuation propagator of the bilocal field $C(x, \tau, \tau')$. Rather than undertake this complex task, in this paper we will limit ourselves to a renormalization group (RG) analysis in powers of $K$ within the large $N_h$ expansion. We will be performing a double expansion in powers of $K$ and $1/N_h$, with $K \sim 1/N_h$.

\begin{figure}
\includegraphics[width=4in]{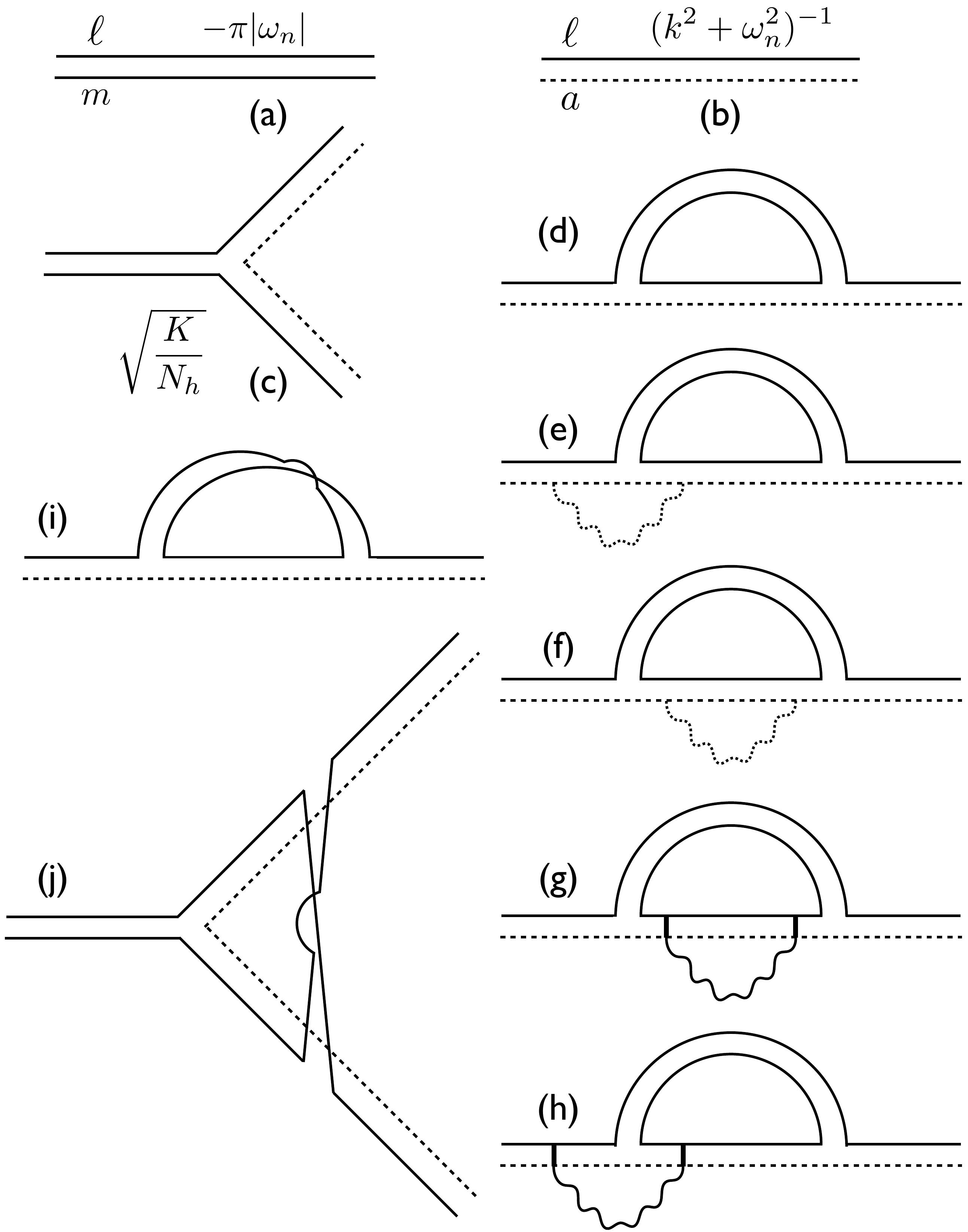}
\caption{Diagrams for the RG computation. (a) Propagator of $D_{\ell m}$.
(b) Propagator for $H_{a \ell}$ (c)~Interaction vertex between $D_{\ell m}$ and $H_{a \ell}$.
(d) Self-energy renormalizations for $H_{a \ell}$ at order $K$. (e-i) Self energy diagrams at order $K/N_h$; the dotted wavy line is the gauge propagator for $A_{a\mu}$, and the dashed wavy lines represent $B_0$ and $B_{1,ab}$ propagators. (j) Vertex renormalization at order $K/N_h$. The vertex renormalization at order $1/N_h$ was computed in Ref.~\onlinecite{SSST19}, and is not shown here. We do not compute diagrams (e-j) in this paper, because they are needed to determine the RG fixed point in (\ref{Kstar}) at order $1/N_h$.} \label{fig:rgdiag}
\end{figure}
To linear order in $K$, the RG equation for $K$ follows from a determination of the scaling dimension of $\mathcal{S}_f$ in (\ref{Sf}); this was already computed in Ref.~\onlinecite{SSST19}, and yields
\beq
\frac{dK}{d \ell} = 2(1 - \Delta_Q) K + \mathcal{O}(K^2) \label{rg1}
\eeq
where $\Delta_Q$ is the scaling dimension of the O($N_h$) order parameter $Q_{\ell m}$ in (\ref{defQ}) at $K=0$; this was computed in Ref.~\onlinecite{SSST19} to be
\beq
\Delta_Q = 1 - \frac{64}{3 \pi^2 N_h} + \mathcal{O}(1/N_h^2) \,. \label{rg2}
\eeq
So we see the $K$ is {\it relevant\/} at large, but finite $N_h$.

The RG analysis is more easily carried out without using bilocal fields. So instead of decoupling $\mathcal{S}_{f}$ in (\ref{Sf}) by the bilocal field $C_{ab}$ in (\ref{Sf+Sb}), we decouple it by a local field $D_{\ell m} (x, \tau)$ \cite{XuChao-Ming};
\beq
\mathcal{S}_f = \frac{1}{2} \int d^2 x d \tau d \tau' \, D_{\ell m} (x, \tau) G_f^{-1} (\tau - \tau')  D_{\ell m} (x, \tau') - \sqrt{\frac{K}{N_h}} \int d^2 x d \tau D_{\ell m} (x, \tau) H_{a \ell} (x, \tau) H_{a m} (x, \tau) \,. \label{Sf2}
\eeq
Here $G_f^{-1}$ is the operator inverse of $G_f (\tau) = 1/\tau^2$. The RG analysis can now be carried out by standard diagrammatic methods, using the Feynman graphs illustrated in Fig.~\ref{fig:rgdiag}.
The RG equation (\ref{rg1}) contains a term of order $K/N_h$. The $H_{a \ell}$ self energy diagram in Fig.~\ref{fig:rgdiag}d contributes to the flow of $K$ at order $K^2$, while the diagrams in Fig.~\ref{fig:rgdiag}e-j contribute the flow of $K$ at order $K^2/N_h$. It will be sufficient for our purposes to only compute the 
diagram in Fig.~\ref{fig:rgdiag}d, which represents the RG implementation of the logarithmic factors discussed in Section~\ref{sec:saddle}. At external frequency $\omega$, and external momentum $p$, we have
\beq
\ref{fig:rgdiag}{\rm d} = 2K \int \frac{d^2 k}{4 \pi^2} \int \frac{d \epsilon}{2 \pi} 
\frac{- \pi |\epsilon|}{[(k+p)^2 + (\epsilon + \omega)^2]} = {\rm constant} - K \omega^2
\int_{\Lambda e^{-\ell}}^\Lambda \frac{d^2k}{4 \pi^2} \frac{1}{k^2}\, \label{fig1d}
\eeq
where $e^{-\ell}$ is the RG rescaling factor.
This self energy can be absorbed into rescalings of $x$, $\tau$, and $H_{al}$ via
\bea
x' &=& x e^{-\ell} \nonumber \\
\tau' &=& \tau e^{-z \ell} \nonumber \\
H_{a \ell}' &=& H_{a \ell} e^{(d+z-2+\eta)\ell/2} \,,
\eea
where $z$ is the dynamic critical exponent and $\eta$ is the anomalous dimension of $H_{a \ell}$.
From (\ref{fig1d}) we obtain
\bea
\eta &=& 0 \nonumber \\
z&=& 1 + \frac{K}{4\pi}\,.
\label{etaz}
\eea
Next, we determine from the first term in (\ref{Sf2}) that the rescaling of $D_{\ell m}$ is
\beq
D_{\ell m}' = D_{\ell m} e^{(1+z)\ell}\,.
\eeq
Finally, from the rescalings of the second term in (\ref{Sf2}), we determine the leading correction to the flow equations in (\ref{rg1}) and (\ref{rg2}) 
\beq
\frac{dK}{d \ell} = \frac{128}{3 \pi^2 } \frac{K}{N_h}  - \frac{K^2}{2\pi} + \mathcal{O}\left(\frac{K}{N_h^2},\frac{K^2}{N_h},K^3\right) \,. \label{rg3}
\eeq

The RG flow equation has an infrared stable fixed point at 
\beq
K^\ast = \frac{256}{3 \pi N_h} + \mathcal{O}\left(\frac{1}{N_h^2} \right). 
\label{Kstar}
\eeq
Note that the relevant direction associated with $g-g_c$ is still present at this fixed point, which is our candidate for cuprate criticality. From (\ref{etaz}) we obtain the dynamic critical exponent
\beq
z = 1 + \frac{64}{3 \pi^2 N_h} + \mathcal{O}\left(\frac{1}{N_h^2} \right)\,.
\label{zres}
\eeq
The field $H_{a \ell}$ is not gauge invariant, and so its anomalous dimension $\eta$ is not well-defined: it can be useful to define a gauge-dependent $\eta$ for intermediate steps in a computation, but it does not directly determine any observable.
In (\ref{etaz}), we obtained $\eta=0$ to leading order in $K$, but we have not explicitly included a wavefunction renormalization from the gauge field. Indeed this wavefunction renormalization is an ingredient \cite{SSST19} in the computation of the anomalous dimension of the gauge-invariant composite operator $Q_{\ell m}$ in (\ref{rg2}), which entered our RG flow equation (\ref{rg3}).

We turn to the critical behavior of the free energy density. In a theory obeying hyperscaling, we expect $F \sim T^{(d+z)/z} = T^{1+ 2/z}$. Using the value of $z$ in (\ref{etaz}), and expanding in powers of $K$, we obtain $F \sim T^3 [ 1 + (K/(2 \pi)) \ln(\bullet) + \ldots]$. We see that this hyperscaling contribution to the free energy perfectly explains the $T^3 \ln (\bullet)$ terms in $F_2$ and $F_3$ in Appendix~\ref{app:linearK}.

\section{Conclusions}
\label{sec:conc}

We have analyzed a model of optimal doping criticality in the cuprates \cite{SSST19,SPSS20}. 
The underlying transition is a Higgs-confinement transition in a SU(2) gauge theory, with the Higgs field corresponding to the spin density wave order in a rotating frame of reference. The Higgs field transforms as an adjoint of the emergent SU(2) gauge field, and so is not directly observable. However, gauge-invariant composites of the Higgs field can break symmetries associated with charge density wave, Ising-nematic, and time-reversal odd scalar spin chirality orders. So the underdoped regime, which corresponds to the Higgs phase, can display one of these orders. In addition, the Higgs condensate need not break the SU(2) gauge symmetry completely, and any unbroken discrete gauge symmetries can lead to bulk topological order with anyonic excitations. The confining phase of the SU(2) gauge theory corresponds to the Fermi liquid in the overdoped regime of the cuprates.

A particularly difficult issue in the treatment of cuprate criticality is the role of the fermions carrying the electromagnetic charge. In many models, these fermions are fractionalized, and also carry emergent gauge charges: then there 
is a singular renormalization of the fermionic excitations at the Fermi surface, which is difficult to treat in a controlled manner. 
In the model of Ref.~\cite{SSST19} (and also in some earlier models \cite{Morinari02,Zaanen02A,Zaanen02B,Mross12,Mross12b}), the electromagnetically charged fermions are argued to be electron-like and have a large Fermi surface (whose volume is given by the conventional Luttinger value). We have shown here that a $1/N_h$ expansion allows a controlled treatment of the consequences of such a Fermi surface. It leads to a quantum field theory which is bilocal in time, with a strongly-coupled fixed point with dynamic critical exponent $z>1$.

We showed that the critical free energy obeyed hyperscaling. At intermediate stages in our computation, hyperscaling violating terms do appear; however we showed in Section~\ref{sec:freeen} and Appendix~\ref{app:gc} and \ref{app:linearK} that such terms cancel after accounting for fluctuation corrections to the position of the quantum critical point. The resulting specific heat is described by (\ref{Cscale}), with a smooth background linear in $T$ specific heat, and a singular hyperscaling preserving contribution. Plots of $C_v/T$ as a function of $T$ and $\Delta$ (the Higgs gap, an energy scale measuring distance from the quantum critical point on the overdoped side) are shown in Fig.~\ref{f:Cv}. 
There is a finite enhancement of the background contribution $\gamma_b$, shown as the first term in (\ref{CDelta}), which can be viewed as an increase in the effective mass of the background fermions from the Higgs fluctuations.
The remaining terms in (\ref{CDelta}), belong to the singular contribution obeying hyperscaling, and show a $\Delta$-dependent finite enhancement in the value $\lim_{T \rightarrow 0} C_v/T$ as the critical point is approached with $\Delta$ becoming smaller. At a fixed $\Delta$, we also found a non-monotonic $T$ dependence in $C_v/T$ at small $\Delta$, with a peak at a value $T \sim \Delta$. This peak is an indication of a crossover associated with the underlying fluctuations of the Higgs field, and could be a useful experimental \cite{Loram01,Michon18,Loram19} diagnostic of the critical point.

\subsection*{Acknowledgements}

We acknowledge useful discussions with Chao-Ming Jian and Cenke Xu.
This research was supported by the National Science Foundation under Grant No.~DMR-2002850.

\appendix 

\section{Position of the critical point}
\label{app:gc}

We work at $T=0$ and $g=g_c$, when $\Delta(T)=0$, and then the saddle point equations in (\ref{s1}) and (\ref{s2}) become
\bea
&~&\int^\Lambda \frac{d^2 k}{4 \pi^2} \int \frac{d \omega}{2 \pi} \frac{1}{k^2 + \omega^2 - 2 \tilde{C} (\omega) + 2 \tilde{C}(0)} = \frac{1}{g_c} \nonumber \\
&~& \tilde{C}(\omega) = \int \frac{d \epsilon}{2 \pi} \tilde{J}_f (\epsilon +\omega) \int^\Lambda \frac{d^2 k}{4 \pi^2}\frac{1}{k^2 + \epsilon^2 - 2 \tilde{C} (\epsilon) + 2 \tilde{C}(0)} \label{saddle}
\eea
We manipulate these equations to obtain the first order correction to $g_c^0 = 4 \pi/\Lambda$. Keeping only the first order term in $K$ in the second equation in (\ref{saddle}) we obtain
\beq
\tilde{C} (\omega) - \tilde{C} (0) = \int \frac{d \epsilon}{2 \pi} \int^\Lambda \frac{d^2 k}{4 \pi^2} \frac{\left[\tilde{J}_f (\epsilon + \omega) - \tilde{J}(\epsilon) \right]}{k^2 + \epsilon^2} \label{saddle1}
\eeq
Inserting this back into the first equation in (\ref{saddle}) we obtain our needed result for $g_c^1$ in (\ref{CSY12}) 
\beq
\frac{1}{g_c^1} = 2  \int \frac{d \omega d \epsilon}{4 \pi^2} \int^\Lambda \frac{d^2 p}{4 \pi^2}
 \int^\Lambda \frac{d^2 k}{4 \pi^2} \frac{\left[\tilde{J}_f (\epsilon + \omega) - \tilde{J}(\epsilon) \right]}{(k^2 + \epsilon^2)(p^2 + \omega^2)^2} \label{gc1}
\eeq

We now evaluate these expressions analytically in the limit $\varkappa \rightarrow 0$. From (\ref{saddle1})
\bea
\tilde{C}(\omega) - \tilde{C} (0) &=& - \pi K \int \frac{d \epsilon}{2 \pi} \int^\Lambda \frac{d^2 k}{4 \pi^2} \frac{\left(|\epsilon + \omega| - |\epsilon| \right)}{k^2 + \epsilon^2} \nonumber \\
&=& - \frac{K}{4}\int \frac{d \epsilon}{2 \pi}\left(|\epsilon + \omega| - |\epsilon| \right) \ln \left(
\frac{\Lambda^2 + \epsilon^2}{\epsilon^2} \right) \nonumber \\
&=& - \frac{K\Lambda \omega}{2 \pi} \tan^{-1} \left( \frac{ \omega}{\Lambda} \right) + \frac{K\Lambda^2}{8 \pi} \ln \left( \frac{ \Lambda^2 + \omega^2}{\Lambda^2} \right) - \frac{K \omega^2}{8 \pi} \ln \left( \frac{ \Lambda^2 + \omega^2}{\omega^2} \right)
\eea
From the first equation in (\ref{saddle}), the value of $g_c$ is then
\beq
\frac{1}{g_c} = \frac{\Lambda}{4\pi} + \frac{1}{2 \pi} \int \frac{d \omega}{2 \pi}
\left( \frac{1}{ \omega^2} - \frac{1}{\Lambda^2 + \omega^2} \right) \left( \tilde{C} (\omega) -\tilde{C} (0) \right)\,.
\eeq
So, evaluating $\omega$ integral
\beq
\frac{1}{g_c^1} = - \frac{K \Lambda}{8 \pi^2} \left( 2 \ln 2 - \frac{1}{2} \right)\,, \quad \mbox{as $\varkappa \rightarrow 0$.} \label{shiftgc}
\eeq

It is interesting to compare the expressions (\ref{gc1}) and (\ref{shiftgc}) with that obtained from the derivative of $F_K$ with respect to $T$ starting from the expression in (\ref{e4}) and (\ref{e4a}). In taking this derivative, we ignore any $T$ dependence that arises from the Matsubara frequency summation: such terms involve derivatives of Bose functions which vanish exponentially at large argument, and so do not contribute to the ultraviolent divergent term $F_K \sim \Lambda [\Delta (T)]^2$ we are interested. 
Furthermore, it is important for our argument that the $T$-dependence of $\Delta(T)$ is compatible with the constraint (\ref{s2}), or more explicitly (\ref{constraint}).
So we obtain
\beq
\frac{1}{3 N_h} \frac{\partial F_K}{\partial T} \approx  \frac{ \partial [\Delta(T)]^2}{\partial T} T^2 \sum_{\omega_n, \epsilon_n} \int^\Lambda \frac{d^2 p}{4 \pi^2}\int^\Lambda \frac{d^2k}{(2\pi)^2} \tilde J_f (\epsilon_n + \omega_n)  \tilde G(k,\epsilon_n)   \tilde G^2 (p,\omega_n) \,. \label{e4b}
\eeq
Now comparing (\ref{e4b}) with (\ref{gc1}) we see that the first term in (\ref{gc1}) has the same form as (\ref{e4b}). The only differences are the frequency integration versus frequency summation, and the presence of the `mass' $[\Delta(T)]^2$ in the Green's function in (\ref{e4b}). However, these differences are not important for the ultraviolet $\Lambda$-dependence we are interested in. The second term in (\ref{gc1}) is needed to cancel the $1/\omega^2$ infrared divergence in the first term. There is no such infrared divergence in (\ref{e4b}) because of the $[\Delta(T)]^2$ mass in $\tilde{G}$. If we were to add a term corresponding to the second term of (\ref{gc1}) to (\ref{e4b}), we would have the concern that this introduces additional ultraviolet divergent terms not in $F_K$. However, this does not happen because
\bea
&~&- \frac{ \partial [\Delta(T)]^2}{\partial T} T^2 \sum_{\omega_n, \epsilon_n} \int^\Lambda \frac{d^2 p}{4 \pi^2}\int^\Lambda \frac{d^2k}{(2\pi)^2} \tilde J_f (\epsilon_n)  \tilde G(k,\epsilon_n)   \tilde G^2 (p,\omega_n) \approx \nonumber \\
&~&~~~~~~~~~~~~~~~~\left[ \sum_{\epsilon_n} \int^\Lambda \frac{d^2k}{(2\pi)^2} \tilde J_f (\epsilon_n)  \tilde G(k,\epsilon_n) \right] \frac{\partial}{\partial T} \left[  \sum_{\omega_n} \int^\Lambda \frac{d^2p}{(2\pi)^2} \tilde G(p,\omega_n) \right]\, \label{gc2}
\eea
vanishes by the constraint equation (\ref{s2}) (up to terms involving derivatives of Bose functions that we are allowed to drop because we are only interested in ultraviolet contributions). Therefore, such a term is not needed, and the correspondence between (\ref{e4b}) and (\ref{gc1}) is complete without it: this explains why the co-efficient of the term divergent as $\sim \Lambda [\Delta(T)]^2 $ in (\ref{FKres}) matches (\ref{FHres}) and (\ref{shiftgc}). The constraint equation (\ref{s2}) was crucial for this argument, as it was in evaluating $F_K$ to obtain (\ref{FKres}) in Appendix~\ref{app:linearK}.

\section{Evaluation of free energy terms proportional to $K$}
\label{app:linearK}

This appendix describes the evaluation of the integrals in (\ref{e7}).
We will split (\ref{e7}) into various contributions, and take the limit $\kappa \rightarrow 0$
\beq
F_K = F_1 + F_2 +F_3
\eeq

The $F_1$ contribution arises from the first two terms in (\ref{e7}); expanding in $\varkappa$ and using (\ref{CSY11a},\ref{CSY14}) we obtain
\bea
\frac{F_1 }{3 N_h} &=&  - \frac{K}{32\pi \varkappa} \int_{\Delta (T)}^{\sqrt{\Lambda^2 + [\Delta(T)]^2}} da \int_{\Delta (T)}^{\sqrt{\Lambda^2 + [\Delta(T)]^2}} db (1 + 2 n(a))(1+2 n(b)) + \mathcal{O}(\varkappa) \nonumber \\
&=& - \frac{\pi K}{2 \varkappa g_R^2}+ \mathcal{O}(\varkappa)
\eea
Using (\ref{CSY11a}) we can write
\beq
\frac{\Delta}{4 \pi} - \frac{\sqrt{\Lambda^2+ \Delta^2} - \Lambda}{4 \pi} = \frac{\Delta}{4 \pi} - \frac{\Delta^2}{8 \pi \Lambda} +\ldots = \frac{1}{g_c^{0}} - \frac{1}{g_R}\,, \label{Deltadef}
\eeq
and so 
\beq
\frac{F_1 }{3 N_h} = - \frac{ K}{32 \pi \varkappa} \left(\sqrt{\Lambda^2+ \Delta^2} - \Delta \right)^2+ \mathcal{O}(\varkappa)
\eeq
So $F_1$ is $T$-independent, and a smooth function of $\Delta$.

The $F_2$ contribution arises from the last two terms in (\ref{e7}), but without the $n(\Omega)$ factor. Performing the $\Omega$ integral, we obtain
\bea
\frac{F_2}{3 N_h}
 &=& - \frac{K}{16\pi^2} \int_{\Delta (T)}^{\sqrt{\Lambda^2 + [\Delta(T)]^2}} da \int_{\Delta (T)}^{\sqrt{\Lambda^2 + [\Delta(T)]^2}} db \Biggl[  \ln (\varkappa) \left[ a (1 + 2 n(b)) + b (1 + 2 n(a)) \right] \nonumber \\
 &~&~~~~~~+ \ln(|a-b|) (b-a)(n(a)-n(b)) +  \ln(|a+b|) (b+a)(1+n(a)+n(b)) \Biggr] \nonumber \\
 &=& - \frac{K \Lambda^2}{4 \pi g_R} \ln (\varkappa) - \frac{K}{16\pi^2} \int_{\Delta (T)}^{\sqrt{\Lambda^2 + [\Delta(T)]^2}} da \int_{\Delta (T)}^{\sqrt{\Lambda^2 + [\Delta(T)]^2}} db \Biggl[ \nonumber \\
 &~&~~~~~~+ \ln(|a-b|) (b-a)(n(a)-n(b)) +  \ln(|a+b|) (b+a)(1+n(a)+n(b)) \Biggr] \label{ef2}
\eea

Finally, the $F_3$ contribution arises from the terms containing the $n(\Omega)$ factor in the last two terms in (\ref{e7})
\bea
\frac{F_3}{3 N_h} &=& - \frac{1}{16\pi^2} \int_{\Delta (T)}^{\sqrt{\Lambda^2 + [\Delta(T)]^2}} da \int_{\Delta (T)}^{\sqrt{\Lambda^2 + [\Delta(T)]^2}} db \Biggl[
\nonumber \\
&~&~~~~~~~+ 2 K \int_0^{\infty}  d \Omega \, \Omega n(\Omega) \, \mathcal{P} \left(\frac{1}{(a-b)^2 - \Omega^2} \right) (b-a)(n(a)-n(b)) \nonumber \\
&~&~~~~~~~+ 2 K \int_0^{\infty} d \Omega \, \Omega n(\Omega) \, \mathcal{P} \left(\frac{1}{(a+b)^2 - \Omega^2} \right)  (a+b) (1+ n(a)+n(b))  \Biggr] \,.
\label{ef3}
\eea

So far, the manipulations have been exact.
Now we will evaluate the integrals while dropping terms which scale as $T^3 \ln (\bullet)$ and are exponentially small in the regime $T \ll \Delta$.

The case of $F_3$ is simpler, so we consider it first. 
In the stated approximation, the only significant contribution in (\ref{ef3}) is
\bea
\frac{F_3}{3 N_h} &\approx& - \frac{K}{8\pi^2} \left[ \int_0^{\infty} d \Omega \, \Omega n(\Omega) \right]  \int_{\Delta}^{\sqrt{\Lambda^2 + \Delta^2}} da \int_{\Delta}^{\sqrt{\Lambda^2 + \Delta^2}} db \, \frac{1}{a+b} \nonumber \\
&=& - \frac{K T^2}{24} \left[ \Lambda \ln(2) -  \Delta \ln \left( \frac{\Lambda}{2\Delta} \right) - \Delta \right]
\label{e20}
\eea

Finally, let us turn to the evaluation of $F_2$. After interchanging the $a$ and $b$ integrands in (\ref{ef2}) so that all the Bose functions are $n(a)$, the $b$ integration can be performed exactly, and we obtain
\bea
 \frac{F_2}{3 N_h}
&=& - \frac{K \Lambda^2}{4 \pi g_R} \ln (\varkappa) - \frac{K}{16\pi^2} \int_{\Delta (T)}^{\sqrt{\Lambda^2 + [\Delta(T)]^2}} da  \Biggl[ - n(a) \left(  a - \Delta(T) \right)^2 \left(\ln \left( a - \Delta (T) \right) - \frac{1}{2} \right) \nonumber \\
 &~&~~~~~~\, + n(a) \left( \sqrt{\Lambda^2 + [\Delta(T)]^2} - a \right)^2 \left(\ln \left( \sqrt{\Lambda^2 + [\Delta(T)]^2} - a \right) - \frac{1}{2} \right)  \nonumber \\
 &~&~~~~~~+ (n(a)+1/2) \left( \sqrt{\Lambda^2 + [\Delta(T)]^2} + a \right)^2 \left(\ln \left( \sqrt{\Lambda^2 + [\Delta(T)]^2} + a \right) - \frac{1}{2} \right)  \nonumber \\
&~&~~~~~~- (n(a)+1/2) \left(  a + \Delta(T) \right)^2 \left(\ln \left( a + \Delta (T) \right) - \frac{1}{2} \right)  \Biggr]\,.
\eea
Now we make the approximation described above, of dropping terms which scale as $T^3 \ln (\bullet)$ and are exponentially small for $T \ll \Delta$; then some of the integrals can be evaluated:
\bea
\frac{F_2}{3 N_h}
 &\approx & - \frac{K \Lambda^2}{4\pi g_R} \ln (\varkappa) - \frac{K}{16 \pi^2} \Lambda^2 (2 \ln \Lambda - 1) \int_{\Delta (T)}^{\infty} da \, n(a)   - \frac{K}{16\pi^2} \int_{\Delta (T)}^{\sqrt{\Lambda^2 + [\Delta(T)]^2}} da  \Biggl[ \nonumber \\
&~&~~~~~~\, + (1/2) \left( \sqrt{\Lambda^2 + [\Delta(T)]^2} + a \right)^2 \left(\ln \left( \sqrt{\Lambda^2 + [\Delta(T)]^2} + a \right) - \frac{1}{2} \right)  \nonumber \\
&~&~~~~~~- (1/2) \left(  a + \Delta(T) \right)^2 \left(\ln \left( a + \Delta (T) \right) - \frac{1}{2} \right)  \Biggr] \nonumber \\
 &=& - \frac{K \Lambda^3}{16 \pi^2} \left[ \ln (\varkappa \Lambda) + \frac{4}{3} \ln 2 - \frac{5}{6} \right] + \frac{K \Lambda^2 \Delta}{16 \pi^2}\left( 1 - \frac{\Delta}{2 \Lambda} \right) \ln (\varkappa )
 - \frac{K}{16\pi^2} \Biggl[ - \frac{ \Delta(T) \Lambda^2}{2} ( 2 \ln \Lambda -1) 
 \nonumber \\
&~&~~~~~~~ + \frac{[\Delta(T)]^2 \Lambda}{4} \left( - 3 + 8 \ln 2 + 2 \ln \Lambda \right) +  \Lambda^2 (2 \ln \Lambda - 1) \int_{\Delta (T)}^{\infty} da \, n(a)   
\Biggr] \,. \label{e77}
\eea
In (\ref{e77}), we have used an expression for $g_R$ that follow from the constraint equations (\ref{CSY11a},\ref{CSY14}) which we write as
\beq
\int_{\Delta(T)}^{\sqrt{\Lambda^2 + [\Delta(T)]^2}} da (1 + 2 n(a)) = \frac{4 \pi}{g_R} \,. \label{constraint}
\eeq
We also used (\ref{Deltadef}) to express $g_R$ in terms $\Lambda$ and $\Delta$. 
We can also use (\ref{constraint}) and (\ref{Deltadef}) to write 
\beq
\int_{\Delta(T)}^{\infty} da \, n(a) =  \frac{\Delta(T)-\Delta}{2} - \frac{[\Delta(T)]^2 - \Delta^2}{4 \Lambda} + \ldots  \,. \label{intn}
\eeq
Now inserting (\ref{intn}) in (\ref{e77}), we obtain
\bea
\frac{F_2}{3 N_h}
 &\approx&  - \frac{K \Lambda^3}{16 \pi^2} \left[ \ln (\varkappa \Lambda) + \frac{4}{3} \ln 2 - \frac{5}{6} \right] + \frac{K \Lambda^2 \Delta}{16 \pi^2}\left( 1 - \frac{\Delta}{2 \Lambda} \right) \left[\ln (\varkappa \Lambda ) - \frac{1}{2} \right] \nonumber \\
&~&~~ - \frac{K \Lambda [\Delta(T)]^2}{16\pi^2} \Biggl[ 2 \ln 2 - \frac{1}{2} \Biggr]\,. \label{F2res}
\eea
It is notable that all terms of order $\Delta \Lambda^2 \ln \Lambda$, $\Delta \Lambda^2$, and $\Delta^2 \Lambda \ln \Lambda$ cancel, even though they appear at intermediate orders. 
This is related to use of the constraint (\ref{s2}), and the discussion in the latter part of Appendix~\ref{app:gc}.


\begin{thebibliography}{33}%
\makeatletter
\providecommand \@ifxundefined [1]{%
 \@ifx{#1\undefined}
}%
\providecommand \@ifnum [1]{%
 \ifnum #1\expandafter \@firstoftwo
 \else \expandafter \@secondoftwo
 \fi
}%
\providecommand \@ifx [1]{%
 \ifx #1\expandafter \@firstoftwo
 \else \expandafter \@secondoftwo
 \fi
}%
\providecommand \natexlab [1]{#1}%
\providecommand \enquote  [1]{``#1''}%
\providecommand \bibnamefont  [1]{#1}%
\providecommand \bibfnamefont [1]{#1}%
\providecommand \citenamefont [1]{#1}%
\providecommand \href@noop [0]{\@secondoftwo}%
\providecommand \href [0]{\begingroup \@sanitize@url \@href}%
\providecommand \@href[1]{\@@startlink{#1}\@@href}%
\providecommand \@@href[1]{\endgroup#1\@@endlink}%
\providecommand \@sanitize@url [0]{\catcode `\\12\catcode `\$12\catcode
  `\&12\catcode `\#12\catcode `\^12\catcode `\_12\catcode `\%12\relax}%
\providecommand \@@startlink[1]{}%
\providecommand \@@endlink[0]{}%
\providecommand \url  [0]{\begingroup\@sanitize@url \@url }%
\providecommand \@url [1]{\endgroup\@href {#1}{\urlprefix }}%
\providecommand \urlprefix  [0]{URL }%
\providecommand \Eprint [0]{\href }%
\providecommand \doibase [0]{http://dx.doi.org/}%
\providecommand \selectlanguage [0]{\@gobble}%
\providecommand \bibinfo  [0]{\@secondoftwo}%
\providecommand \bibfield  [0]{\@secondoftwo}%
\providecommand \translation [1]{[#1]}%
\providecommand \BibitemOpen [0]{}%
\providecommand \bibitemStop [0]{}%
\providecommand \bibitemNoStop [0]{.\EOS\space}%
\providecommand \EOS [0]{\spacefactor3000\relax}%
\providecommand \BibitemShut  [1]{\csname bibitem#1\endcsname}%
\let\auto@bib@innerbib\@empty
\bibitem [{\citenamefont {{Proust}}\ and\ \citenamefont
  {{Taillefer}}(2019)}]{CPLT18}%
  \BibitemOpen
  \bibfield  {author} {\bibinfo {author} {\bibfnamefont {C.}~\bibnamefont
  {{Proust}}}\ and\ \bibinfo {author} {\bibfnamefont {L.}~\bibnamefont
  {{Taillefer}}},\ }\bibfield  {title} {\enquote {\bibinfo {title} {{The
  Remarkable Underlying Ground States of Cuprate Superconductors}},}\ }\href
  {\doibase 10.1146/annurev-conmatphys-031218-013210} {\bibfield  {journal}
  {\bibinfo  {journal} {Annual Review of Condensed Matter Physics}\ }\textbf
  {\bibinfo {volume} {10}},\ \bibinfo {pages} {409} (\bibinfo {year} {2019})},\
  \Eprint {http://arxiv.org/abs/1807.05074} {arXiv:1807.05074
  [cond-mat.supr-con]} \BibitemShut {NoStop}%
\bibitem [{\citenamefont {{Vishik}}\ \emph {et~al.}(2012)\citenamefont
  {{Vishik}}, \citenamefont {{Hashimoto}}, \citenamefont {{He}}, \citenamefont
  {{Lee}}, \citenamefont {{Schmitt}}, \citenamefont {{Lu}}, \citenamefont
  {{Moore}}, \citenamefont {{Zhang}}, \citenamefont {{Meevasana}},
  \citenamefont {{Sasagawa}}, \citenamefont {{Uchida}}, \citenamefont
  {{Fujita}}, \citenamefont {{Ishida}}, \citenamefont {{Ishikado}},
  \citenamefont {{Yoshida}}, \citenamefont {{Eisaki}}, \citenamefont
  {{Hussain}}, \citenamefont {{Devereaux}},\ and\ \citenamefont
  {{Shen}}}]{Vishik2012}%
  \BibitemOpen
  \bibfield  {author} {\bibinfo {author} {\bibfnamefont {I.~M.}\ \bibnamefont
  {{Vishik}}}, \bibinfo {author} {\bibfnamefont {M.}~\bibnamefont
  {{Hashimoto}}}, \bibinfo {author} {\bibfnamefont {R.-H.}\ \bibnamefont
  {{He}}}, \bibinfo {author} {\bibfnamefont {W.-S.}\ \bibnamefont {{Lee}}},
  \bibinfo {author} {\bibfnamefont {F.}~\bibnamefont {{Schmitt}}}, \bibinfo
  {author} {\bibfnamefont {D.}~\bibnamefont {{Lu}}}, \bibinfo {author}
  {\bibfnamefont {R.~G.}\ \bibnamefont {{Moore}}}, \bibinfo {author}
  {\bibfnamefont {C.}~\bibnamefont {{Zhang}}}, \bibinfo {author} {\bibfnamefont
  {W.}~\bibnamefont {{Meevasana}}}, \bibinfo {author} {\bibfnamefont
  {T.}~\bibnamefont {{Sasagawa}}}, \bibinfo {author} {\bibfnamefont
  {S.}~\bibnamefont {{Uchida}}}, \bibinfo {author} {\bibfnamefont
  {K.}~\bibnamefont {{Fujita}}}, \bibinfo {author} {\bibfnamefont
  {S.}~\bibnamefont {{Ishida}}}, \bibinfo {author} {\bibfnamefont
  {M.}~\bibnamefont {{Ishikado}}}, \bibinfo {author} {\bibfnamefont
  {Y.}~\bibnamefont {{Yoshida}}}, \bibinfo {author} {\bibfnamefont
  {H.}~\bibnamefont {{Eisaki}}}, \bibinfo {author} {\bibfnamefont
  {Z.}~\bibnamefont {{Hussain}}}, \bibinfo {author} {\bibfnamefont {T.~P.}\
  \bibnamefont {{Devereaux}}}, \ and\ \bibinfo {author} {\bibfnamefont {Z.-X.}\
  \bibnamefont {{Shen}}},\ }\bibfield  {title} {\enquote {\bibinfo {title}
  {{Phase competition in trisected superconducting dome}},}\ }\href {\doibase
  10.1073/pnas.1209471109} {\bibfield  {journal} {\bibinfo  {journal}
  {Proceedings of the National Academy of Science}\ }\textbf {\bibinfo {volume}
  {109}},\ \bibinfo {pages} {18332} (\bibinfo {year} {2012})},\ \Eprint
  {http://arxiv.org/abs/1209.6514} {arXiv:1209.6514 [cond-mat.supr-con]}
  \BibitemShut {NoStop}%
\bibitem [{\citenamefont {{He}}\ \emph {et~al.}(2014)\citenamefont {{He}},
  \citenamefont {{Yin}}, \citenamefont {{Zech}}, \citenamefont
  {{Soumyanarayanan}}, \citenamefont {{Yee}}, \citenamefont {{Williams}},
  \citenamefont {{Boyer}}, \citenamefont {{Chatterjee}}, \citenamefont
  {{Wise}}, \citenamefont {{Zeljkovic}}, \citenamefont {{Kondo}}, \citenamefont
  {{Takeuchi}}, \citenamefont {{Ikuta}}, \citenamefont {{Mistark}},
  \citenamefont {{Markiewicz}}, \citenamefont {{Bansil}}, \citenamefont
  {{Sachdev}}, \citenamefont {{Hudson}},\ and\ \citenamefont
  {{Hoffman}}}]{He14}%
  \BibitemOpen
  \bibfield  {author} {\bibinfo {author} {\bibfnamefont {Y.}~\bibnamefont
  {{He}}}, \bibinfo {author} {\bibfnamefont {Y.}~\bibnamefont {{Yin}}},
  \bibinfo {author} {\bibfnamefont {M.}~\bibnamefont {{Zech}}}, \bibinfo
  {author} {\bibfnamefont {A.}~\bibnamefont {{Soumyanarayanan}}}, \bibinfo
  {author} {\bibfnamefont {M.~M.}\ \bibnamefont {{Yee}}}, \bibinfo {author}
  {\bibfnamefont {T.}~\bibnamefont {{Williams}}}, \bibinfo {author}
  {\bibfnamefont {M.~C.}\ \bibnamefont {{Boyer}}}, \bibinfo {author}
  {\bibfnamefont {K.}~\bibnamefont {{Chatterjee}}}, \bibinfo {author}
  {\bibfnamefont {W.~D.}\ \bibnamefont {{Wise}}}, \bibinfo {author}
  {\bibfnamefont {I.}~\bibnamefont {{Zeljkovic}}}, \bibinfo {author}
  {\bibfnamefont {T.}~\bibnamefont {{Kondo}}}, \bibinfo {author} {\bibfnamefont
  {T.}~\bibnamefont {{Takeuchi}}}, \bibinfo {author} {\bibfnamefont
  {H.}~\bibnamefont {{Ikuta}}}, \bibinfo {author} {\bibfnamefont
  {P.}~\bibnamefont {{Mistark}}}, \bibinfo {author} {\bibfnamefont {R.~S.}\
  \bibnamefont {{Markiewicz}}}, \bibinfo {author} {\bibfnamefont
  {A.}~\bibnamefont {{Bansil}}}, \bibinfo {author} {\bibfnamefont
  {S.}~\bibnamefont {{Sachdev}}}, \bibinfo {author} {\bibfnamefont {E.~W.}\
  \bibnamefont {{Hudson}}}, \ and\ \bibinfo {author} {\bibfnamefont {J.~E.}\
  \bibnamefont {{Hoffman}}},\ }\bibfield  {title} {\enquote {\bibinfo {title}
  {{Fermi Surface and Pseudogap Evolution in a Cuprate Superconductor}},}\
  }\href {\doibase 10.1126/science.1248221} {\bibfield  {journal} {\bibinfo
  {journal} {Science}\ }\textbf {\bibinfo {volume} {344}},\ \bibinfo {pages}
  {608} (\bibinfo {year} {2014})},\ \Eprint {http://arxiv.org/abs/1305.2778}
  {arXiv:1305.2778 [cond-mat.supr-con]} \BibitemShut {NoStop}%
\bibitem [{\citenamefont {{Fujita}}\ \emph {et~al.}(2014)\citenamefont
  {{Fujita}}, \citenamefont {{Kim}}, \citenamefont {{Lee}}, \citenamefont
  {{Lee}}, \citenamefont {{Hamidian}}, \citenamefont {{Firmo}}, \citenamefont
  {{Mukhopadhyay}}, \citenamefont {{Eisaki}}, \citenamefont {{Uchida}},
  \citenamefont {{Lawler}}, \citenamefont {{Kim}},\ and\ \citenamefont
  {{Davis}}}]{Fujita14}%
  \BibitemOpen
  \bibfield  {author} {\bibinfo {author} {\bibfnamefont {K.}~\bibnamefont
  {{Fujita}}}, \bibinfo {author} {\bibfnamefont {C.~K.}\ \bibnamefont {{Kim}}},
  \bibinfo {author} {\bibfnamefont {I.}~\bibnamefont {{Lee}}}, \bibinfo
  {author} {\bibfnamefont {J.}~\bibnamefont {{Lee}}}, \bibinfo {author}
  {\bibfnamefont {M.~H.}\ \bibnamefont {{Hamidian}}}, \bibinfo {author}
  {\bibfnamefont {I.~A.}\ \bibnamefont {{Firmo}}}, \bibinfo {author}
  {\bibfnamefont {S.}~\bibnamefont {{Mukhopadhyay}}}, \bibinfo {author}
  {\bibfnamefont {H.}~\bibnamefont {{Eisaki}}}, \bibinfo {author}
  {\bibfnamefont {S.}~\bibnamefont {{Uchida}}}, \bibinfo {author}
  {\bibfnamefont {M.~J.}\ \bibnamefont {{Lawler}}}, \bibinfo {author}
  {\bibfnamefont {E.-A.}\ \bibnamefont {{Kim}}}, \ and\ \bibinfo {author}
  {\bibfnamefont {J.~C.}\ \bibnamefont {{Davis}}},\ }\bibfield  {title}
  {\enquote {\bibinfo {title} {{Simultaneous Transitions in Cuprate
  Momentum-Space Topology and Electronic Symmetry Breaking}},}\ }\href
  {\doibase 10.1126/science.1248783} {\bibfield  {journal} {\bibinfo  {journal}
  {Science}\ }\textbf {\bibinfo {volume} {344}},\ \bibinfo {pages} {612}
  (\bibinfo {year} {2014})},\ \Eprint {http://arxiv.org/abs/1403.7788}
  {arXiv:1403.7788 [cond-mat.supr-con]} \BibitemShut {NoStop}%
\bibitem [{\citenamefont {{Badoux}}\ \emph {et~al.}(2016)\citenamefont
  {{Badoux}}, \citenamefont {{Tabis}}, \citenamefont {{Lalibert{\'e}}},
  \citenamefont {{Grissonnanche}}, \citenamefont {{Vignolle}}, \citenamefont
  {{Vignolles}}, \citenamefont {{B{\'e}ard}}, \citenamefont {{Bonn}},
  \citenamefont {{Hardy}}, \citenamefont {{Liang}}, \citenamefont
  {{Doiron-Leyraud}}, \citenamefont {{Taillefer}},\ and\ \citenamefont
  {{Proust}}}]{Badoux16}%
  \BibitemOpen
  \bibfield  {author} {\bibinfo {author} {\bibfnamefont {S.}~\bibnamefont
  {{Badoux}}}, \bibinfo {author} {\bibfnamefont {W.}~\bibnamefont {{Tabis}}},
  \bibinfo {author} {\bibfnamefont {F.}~\bibnamefont {{Lalibert{\'e}}}},
  \bibinfo {author} {\bibfnamefont {G.}~\bibnamefont {{Grissonnanche}}},
  \bibinfo {author} {\bibfnamefont {B.}~\bibnamefont {{Vignolle}}}, \bibinfo
  {author} {\bibfnamefont {D.}~\bibnamefont {{Vignolles}}}, \bibinfo {author}
  {\bibfnamefont {J.}~\bibnamefont {{B{\'e}ard}}}, \bibinfo {author}
  {\bibfnamefont {D.~A.}\ \bibnamefont {{Bonn}}}, \bibinfo {author}
  {\bibfnamefont {W.~N.}\ \bibnamefont {{Hardy}}}, \bibinfo {author}
  {\bibfnamefont {R.}~\bibnamefont {{Liang}}}, \bibinfo {author} {\bibfnamefont
  {N.}~\bibnamefont {{Doiron-Leyraud}}}, \bibinfo {author} {\bibfnamefont
  {L.}~\bibnamefont {{Taillefer}}}, \ and\ \bibinfo {author} {\bibfnamefont
  {C.}~\bibnamefont {{Proust}}},\ }\bibfield  {title} {\enquote {\bibinfo
  {title} {{Change of carrier density at the pseudogap critical point of a
  cuprate superconductor}},}\ }\href {\doibase 10.1038/nature16983} {\bibfield
  {journal} {\bibinfo  {journal} {Nature}\ }\textbf {\bibinfo {volume} {531}},\
  \bibinfo {pages} {210} (\bibinfo {year} {2016})},\ \Eprint
  {http://arxiv.org/abs/1511.08162} {arXiv:1511.08162 [cond-mat.supr-con]}
  \BibitemShut {NoStop}%
\bibitem [{\citenamefont {Loram}\ \emph {et~al.}(2001)\citenamefont {Loram},
  \citenamefont {Luo}, \citenamefont {Cooper}, \citenamefont {Liang},\ and\
  \citenamefont {Tallon}}]{Loram01}%
  \BibitemOpen
  \bibfield  {author} {\bibinfo {author} {\bibfnamefont {J.}~\bibnamefont
  {Loram}}, \bibinfo {author} {\bibfnamefont {J.}~\bibnamefont {Luo}}, \bibinfo
  {author} {\bibfnamefont {J.}~\bibnamefont {Cooper}}, \bibinfo {author}
  {\bibfnamefont {W.}~\bibnamefont {Liang}}, \ and\ \bibinfo {author}
  {\bibfnamefont {J.}~\bibnamefont {Tallon}},\ }\bibfield  {title} {\enquote
  {\bibinfo {title} {Evidence on the pseudogap and condensate from the
  electronic specific heat},}\ }\href {\doibase
  https://doi.org/10.1016/S0022-3697(00)00101-3} {\bibfield  {journal}
  {\bibinfo  {journal} {Journal of Physics and Chemistry of Solids}\ }\textbf
  {\bibinfo {volume} {62}},\ \bibinfo {pages} {59 } (\bibinfo {year}
  {2001})}\BibitemShut {NoStop}%
\bibitem [{\citenamefont {{Michon}}\ \emph {et~al.}(2019)\citenamefont
  {{Michon}}, \citenamefont {{Girod}}, \citenamefont {{Badoux}}, \citenamefont
  {{Ka{\v{c}}mar{\v{c}}{\'\i}k}}, \citenamefont {{Ma}}, \citenamefont
  {{Dragomir}}, \citenamefont {{Dabkowska}}, \citenamefont {{Gaulin}},
  \citenamefont {{Zhou}}, \citenamefont {{Pyon}}, \citenamefont {{Takayama}},
  \citenamefont {{Takagi}}, \citenamefont {{Verret}}, \citenamefont
  {{Doiron-Leyraud}}, \citenamefont {{Marcenat}}, \citenamefont {{Taillefer}},\
  and\ \citenamefont {{Klein}}}]{Michon18}%
  \BibitemOpen
  \bibfield  {author} {\bibinfo {author} {\bibfnamefont {B.}~\bibnamefont
  {{Michon}}}, \bibinfo {author} {\bibfnamefont {C.}~\bibnamefont {{Girod}}},
  \bibinfo {author} {\bibfnamefont {S.}~\bibnamefont {{Badoux}}}, \bibinfo
  {author} {\bibfnamefont {J.}~\bibnamefont {{Ka{\v{c}}mar{\v{c}}{\'\i}k}}},
  \bibinfo {author} {\bibfnamefont {Q.}~\bibnamefont {{Ma}}}, \bibinfo {author}
  {\bibfnamefont {M.}~\bibnamefont {{Dragomir}}}, \bibinfo {author}
  {\bibfnamefont {H.~A.}\ \bibnamefont {{Dabkowska}}}, \bibinfo {author}
  {\bibfnamefont {B.~D.}\ \bibnamefont {{Gaulin}}}, \bibinfo {author}
  {\bibfnamefont {J.~S.}\ \bibnamefont {{Zhou}}}, \bibinfo {author}
  {\bibfnamefont {S.}~\bibnamefont {{Pyon}}}, \bibinfo {author} {\bibfnamefont
  {T.}~\bibnamefont {{Takayama}}}, \bibinfo {author} {\bibfnamefont
  {H.}~\bibnamefont {{Takagi}}}, \bibinfo {author} {\bibfnamefont
  {S.}~\bibnamefont {{Verret}}}, \bibinfo {author} {\bibfnamefont
  {N.}~\bibnamefont {{Doiron-Leyraud}}}, \bibinfo {author} {\bibfnamefont
  {C.}~\bibnamefont {{Marcenat}}}, \bibinfo {author} {\bibfnamefont
  {L.}~\bibnamefont {{Taillefer}}}, \ and\ \bibinfo {author} {\bibfnamefont
  {T.}~\bibnamefont {{Klein}}},\ }\bibfield  {title} {\enquote {\bibinfo
  {title} {{Thermodynamic signatures of quantum criticality in cuprate
  superconductors}},}\ }\href {\doibase 10.1038/s41586-019-0932-x} {\bibfield
  {journal} {\bibinfo  {journal} {Nature}\ }\textbf {\bibinfo {volume} {567}},\
  \bibinfo {pages} {218} (\bibinfo {year} {2019})},\ \Eprint
  {http://arxiv.org/abs/1804.08502} {arXiv:1804.08502 [cond-mat.supr-con]}
  \BibitemShut {NoStop}%
\bibitem [{\citenamefont {{Tallon}}\ \emph {et~al.}(2019)\citenamefont
  {{Tallon}}, \citenamefont {{Storey}}, \citenamefont {{Cooper}},\ and\
  \citenamefont {{Loram}}}]{Loram19}%
  \BibitemOpen
  \bibfield  {author} {\bibinfo {author} {\bibfnamefont {J.~L.}\ \bibnamefont
  {{Tallon}}}, \bibinfo {author} {\bibfnamefont {J.~G.}\ \bibnamefont
  {{Storey}}}, \bibinfo {author} {\bibfnamefont {J.~R.}\ \bibnamefont
  {{Cooper}}}, \ and\ \bibinfo {author} {\bibfnamefont {J.~W.}\ \bibnamefont
  {{Loram}}},\ }\bibfield  {title} {\enquote {\bibinfo {title} {{Locating the
  pseudogap closing point in cuprate superconductors: absence of entrant or
  reentrant behavior}},}\ }\href@noop {} {\  (\bibinfo {year} {2019})},\
  \Eprint {http://arxiv.org/abs/1907.12018} {arXiv:1907.12018
  [cond-mat.supr-con]} \BibitemShut {NoStop}%
\bibitem [{\citenamefont {{Tang}}\ \emph {et~al.}(2018)\citenamefont {{Tang}},
  \citenamefont {{Mangin-Thro}}, \citenamefont {{Wildes}}, \citenamefont
  {{Chan}}, \citenamefont {{Dorow}}, \citenamefont {{Jeong}}, \citenamefont
  {{Sidis}}, \citenamefont {{Greven}},\ and\ \citenamefont
  {{Bourges}}}]{Bourges18}%
  \BibitemOpen
  \bibfield  {author} {\bibinfo {author} {\bibfnamefont {Y.}~\bibnamefont
  {{Tang}}}, \bibinfo {author} {\bibfnamefont {L.}~\bibnamefont
  {{Mangin-Thro}}}, \bibinfo {author} {\bibfnamefont {A.}~\bibnamefont
  {{Wildes}}}, \bibinfo {author} {\bibfnamefont {M.~K.}\ \bibnamefont
  {{Chan}}}, \bibinfo {author} {\bibfnamefont {C.~J.}\ \bibnamefont {{Dorow}}},
  \bibinfo {author} {\bibfnamefont {J.}~\bibnamefont {{Jeong}}}, \bibinfo
  {author} {\bibfnamefont {Y.}~\bibnamefont {{Sidis}}}, \bibinfo {author}
  {\bibfnamefont {M.}~\bibnamefont {{Greven}}}, \ and\ \bibinfo {author}
  {\bibfnamefont {P.}~\bibnamefont {{Bourges}}},\ }\bibfield  {title} {\enquote
  {\bibinfo {title} {{Orientation of the intra-unit-cell magnetic moment in the
  high-$T_{c}$ superconductor HgBa$_{2}$CuO$_{4 +{\ensuremath{\delta}}}$}},}\
  }\href {\doibase 10.1103/PhysRevB.98.214418} {\bibfield  {journal} {\bibinfo
  {journal} {\prb}\ }\textbf {\bibinfo {volume} {98}},\ \bibinfo {eid} {214418}
  (\bibinfo {year} {2018})},\ \Eprint {http://arxiv.org/abs/1805.02063}
  {arXiv:1805.02063 [cond-mat.supr-con]} \BibitemShut {NoStop}%
\bibitem [{\citenamefont {Chen}\ \emph {et~al.}(2019)\citenamefont {Chen},
  \citenamefont {Hashimoto}, \citenamefont {He}, \citenamefont {Song},
  \citenamefont {Xu}, \citenamefont {He}, \citenamefont {Devereaux},
  \citenamefont {Eisaki}, \citenamefont {Lu}, \citenamefont {Zaanen},\ and\
  \citenamefont {Shen}}]{Shen19}%
  \BibitemOpen
  \bibfield  {author} {\bibinfo {author} {\bibfnamefont {S.-D.}\ \bibnamefont
  {Chen}}, \bibinfo {author} {\bibfnamefont {M.}~\bibnamefont {Hashimoto}},
  \bibinfo {author} {\bibfnamefont {Y.}~\bibnamefont {He}}, \bibinfo {author}
  {\bibfnamefont {D.}~\bibnamefont {Song}}, \bibinfo {author} {\bibfnamefont
  {K.-J.}\ \bibnamefont {Xu}}, \bibinfo {author} {\bibfnamefont {J.-F.}\
  \bibnamefont {He}}, \bibinfo {author} {\bibfnamefont {T.~P.}\ \bibnamefont
  {Devereaux}}, \bibinfo {author} {\bibfnamefont {H.}~\bibnamefont {Eisaki}},
  \bibinfo {author} {\bibfnamefont {D.-H.}\ \bibnamefont {Lu}}, \bibinfo
  {author} {\bibfnamefont {J.}~\bibnamefont {Zaanen}}, \ and\ \bibinfo {author}
  {\bibfnamefont {Z.-X.}\ \bibnamefont {Shen}},\ }\bibfield  {title} {\enquote
  {\bibinfo {title} {{Incoherent strange metal sharply bounded by a critical
  doping in Bi2212}},}\ }\href {\doibase 10.1126/science.aaw8850} {\bibfield
  {journal} {\bibinfo  {journal} {Science}\ }\textbf {\bibinfo {volume}
  {366}},\ \bibinfo {pages} {1099} (\bibinfo {year} {2019})}\BibitemShut
  {NoStop}%
\bibitem [{\citenamefont {{Panagopoulos}}\ \emph {et~al.}(2002)\citenamefont
  {{Panagopoulos}}, \citenamefont {{Tallon}}, \citenamefont {{Rainford}},
  \citenamefont {{Xiang}}, \citenamefont {{Cooper}},\ and\ \citenamefont
  {{Scott}}}]{CPana1}%
  \BibitemOpen
  \bibfield  {author} {\bibinfo {author} {\bibfnamefont {C.}~\bibnamefont
  {{Panagopoulos}}}, \bibinfo {author} {\bibfnamefont {J.~L.}\ \bibnamefont
  {{Tallon}}}, \bibinfo {author} {\bibfnamefont {B.~D.}\ \bibnamefont
  {{Rainford}}}, \bibinfo {author} {\bibfnamefont {T.}~\bibnamefont {{Xiang}}},
  \bibinfo {author} {\bibfnamefont {J.~R.}\ \bibnamefont {{Cooper}}}, \ and\
  \bibinfo {author} {\bibfnamefont {C.~A.}\ \bibnamefont {{Scott}}},\
  }\bibfield  {title} {\enquote {\bibinfo {title} {{Evidence for a generic
  quantum transition in high-$T_{c}$ cuprates}},}\ }\href {\doibase
  10.1103/PhysRevB.66.064501} {\bibfield  {journal} {\bibinfo  {journal} {Phys.
  Rev. B}\ }\textbf {\bibinfo {volume} {66}},\ \bibinfo {eid} {064501}
  (\bibinfo {year} {2002})},\ \Eprint {http://arxiv.org/abs/cond-mat/0204106}
  {arXiv:cond-mat/0204106 [cond-mat.supr-con]} \BibitemShut {NoStop}%
\bibitem [{\citenamefont {{Panagopoulos}}\ \emph {et~al.}(2004)\citenamefont
  {{Panagopoulos}}, \citenamefont {{Petrovic}}, \citenamefont {{Hillier}},
  \citenamefont {{Tallon}}, \citenamefont {{Scott}},\ and\ \citenamefont
  {{Rainford}}}]{CPana2}%
  \BibitemOpen
  \bibfield  {author} {\bibinfo {author} {\bibfnamefont {C.}~\bibnamefont
  {{Panagopoulos}}}, \bibinfo {author} {\bibfnamefont {A.~P.}\ \bibnamefont
  {{Petrovic}}}, \bibinfo {author} {\bibfnamefont {A.~D.}\ \bibnamefont
  {{Hillier}}}, \bibinfo {author} {\bibfnamefont {J.~L.}\ \bibnamefont
  {{Tallon}}}, \bibinfo {author} {\bibfnamefont {C.~A.}\ \bibnamefont
  {{Scott}}}, \ and\ \bibinfo {author} {\bibfnamefont {B.~D.}\ \bibnamefont
  {{Rainford}}},\ }\bibfield  {title} {\enquote {\bibinfo {title} {{Exposing
  the spin-glass ground state of the nonsuperconducting
  La$_{2-x}$Sr$_{x}$Cu$_{1-y}$Zn$_{y}$O$_{4}$ high-$T_{c}$ oxide}},}\ }\href
  {\doibase 10.1103/PhysRevB.69.144510} {\bibfield  {journal} {\bibinfo
  {journal} {Phys. Rev. B}\ }\textbf {\bibinfo {volume} {69}},\ \bibinfo {eid}
  {144510} (\bibinfo {year} {2004})},\ \Eprint
  {http://arxiv.org/abs/cond-mat/0307392} {arXiv:cond-mat/0307392
  [cond-mat.supr-con]} \BibitemShut {NoStop}%
\bibitem [{\citenamefont {{Frachet}}\ \emph {et~al.}(2019)\citenamefont
  {{Frachet}}, \citenamefont {{Vinograd}}, \citenamefont {{Zhou}},
  \citenamefont {{Benhabib}}, \citenamefont {{Wu}}, \citenamefont {{Mayaffre}},
  \citenamefont {{Kr{\"a}mer}}, \citenamefont {{Ramakrishna}}, \citenamefont
  {{Reyes}}, \citenamefont {{Debray}}, \citenamefont {{Kurosawa}},
  \citenamefont {{Momono}}, \citenamefont {{Oda}}, \citenamefont {{Komiya}},
  \citenamefont {{Ono}}, \citenamefont {{Horio}}, \citenamefont {{Chang}},
  \citenamefont {{Proust}}, \citenamefont {{LeBoeuf}},\ and\ \citenamefont
  {{Julien}}}]{Julien19}%
  \BibitemOpen
  \bibfield  {author} {\bibinfo {author} {\bibfnamefont {M.}~\bibnamefont
  {{Frachet}}}, \bibinfo {author} {\bibfnamefont {I.}~\bibnamefont
  {{Vinograd}}}, \bibinfo {author} {\bibfnamefont {R.}~\bibnamefont {{Zhou}}},
  \bibinfo {author} {\bibfnamefont {S.}~\bibnamefont {{Benhabib}}}, \bibinfo
  {author} {\bibfnamefont {S.}~\bibnamefont {{Wu}}}, \bibinfo {author}
  {\bibfnamefont {H.}~\bibnamefont {{Mayaffre}}}, \bibinfo {author}
  {\bibfnamefont {S.}~\bibnamefont {{Kr{\"a}mer}}}, \bibinfo {author}
  {\bibfnamefont {S.~K.}\ \bibnamefont {{Ramakrishna}}}, \bibinfo {author}
  {\bibfnamefont {A.}~\bibnamefont {{Reyes}}}, \bibinfo {author} {\bibfnamefont
  {J.}~\bibnamefont {{Debray}}}, \bibinfo {author} {\bibfnamefont
  {T.}~\bibnamefont {{Kurosawa}}}, \bibinfo {author} {\bibfnamefont
  {N.}~\bibnamefont {{Momono}}}, \bibinfo {author} {\bibfnamefont
  {M.}~\bibnamefont {{Oda}}}, \bibinfo {author} {\bibfnamefont
  {S.}~\bibnamefont {{Komiya}}}, \bibinfo {author} {\bibfnamefont
  {S.}~\bibnamefont {{Ono}}}, \bibinfo {author} {\bibfnamefont
  {M.}~\bibnamefont {{Horio}}}, \bibinfo {author} {\bibfnamefont
  {J.}~\bibnamefont {{Chang}}}, \bibinfo {author} {\bibfnamefont
  {C.}~\bibnamefont {{Proust}}}, \bibinfo {author} {\bibfnamefont
  {D.}~\bibnamefont {{LeBoeuf}}}, \ and\ \bibinfo {author} {\bibfnamefont
  {M.-H.}\ \bibnamefont {{Julien}}},\ }\bibfield  {title} {\enquote {\bibinfo
  {title} {{Hidden magnetism at the pseudogap critical point of a high
  temperature superconductor}},}\ }\href@noop {} {\  (\bibinfo {year}
  {2019})},\ \Eprint {http://arxiv.org/abs/1909.10258} {arXiv:1909.10258
  [cond-mat.supr-con]} \BibitemShut {NoStop}%
\bibitem [{\citenamefont {Sachdev}\ \emph {et~al.}(2019)\citenamefont
  {Sachdev}, \citenamefont {Scammell}, \citenamefont {Scheurer},\ and\
  \citenamefont {Tarnopolsky}}]{SSST19}%
  \BibitemOpen
  \bibfield  {author} {\bibinfo {author} {\bibfnamefont {S.}~\bibnamefont
  {Sachdev}}, \bibinfo {author} {\bibfnamefont {H.~D.}\ \bibnamefont
  {Scammell}}, \bibinfo {author} {\bibfnamefont {M.~S.}\ \bibnamefont
  {Scheurer}}, \ and\ \bibinfo {author} {\bibfnamefont {G.}~\bibnamefont
  {Tarnopolsky}},\ }\bibfield  {title} {\enquote {\bibinfo {title} {{Gauge
  theory for the cuprates near optimal doping}},}\ }\href {\doibase
  10.1103/PhysRevB.99.054516} {\bibfield  {journal} {\bibinfo  {journal} {Phys.
  Rev. B}\ }\textbf {\bibinfo {volume} {99}},\ \bibinfo {pages} {054516}
  (\bibinfo {year} {2019})},\ \Eprint {http://arxiv.org/abs/1811.04930}
  {arXiv:1811.04930 [cond-mat.str-el]} \BibitemShut {NoStop}%
\bibitem [{\citenamefont {Scammell}\ \emph {et~al.}(2020)\citenamefont
  {Scammell}, \citenamefont {Patekar}, \citenamefont {Scheurer},\ and\
  \citenamefont {Sachdev}}]{SPSS20}%
  \BibitemOpen
  \bibfield  {author} {\bibinfo {author} {\bibfnamefont {H.~D.}\ \bibnamefont
  {Scammell}}, \bibinfo {author} {\bibfnamefont {K.}~\bibnamefont {Patekar}},
  \bibinfo {author} {\bibfnamefont {M.~S.}\ \bibnamefont {Scheurer}}, \ and\
  \bibinfo {author} {\bibfnamefont {S.}~\bibnamefont {Sachdev}},\ }\bibfield
  {title} {\enquote {\bibinfo {title} {{Phases of SU(2) gauge theory with
  multiple adjoint Higgs fields in 2+1 dimensions}},}\ }\href {\doibase
  10.1103/PhysRevB.101.205124} {\bibfield  {journal} {\bibinfo  {journal}
  {Phys. Rev. B}\ }\textbf {\bibinfo {volume} {101}},\ \bibinfo {pages}
  {205124} (\bibinfo {year} {2020})},\ \Eprint
  {http://arxiv.org/abs/1912.06108} {arXiv:1912.06108 [cond-mat.str-el]}
  \BibitemShut {NoStop}%
\bibitem [{\citenamefont {{Sachdev}}\ and\ \citenamefont
  {{Morinari}}(2002)}]{Morinari02}%
  \BibitemOpen
  \bibfield  {author} {\bibinfo {author} {\bibfnamefont {S.}~\bibnamefont
  {{Sachdev}}}\ and\ \bibinfo {author} {\bibfnamefont {T.}~\bibnamefont
  {{Morinari}}},\ }\bibfield  {title} {\enquote {\bibinfo {title} {{Strongly
  coupled quantum criticality with a Fermi surface in two dimensions:
  Fractionalization of spin and charge collective modes}},}\ }\href {\doibase
  10.1103/PhysRevB.66.235117} {\bibfield  {journal} {\bibinfo  {journal} {Phys.
  Rev. B}\ }\textbf {\bibinfo {volume} {66}},\ \bibinfo {eid} {235117}
  (\bibinfo {year} {2002})},\ \Eprint {http://arxiv.org/abs/cond-mat/0207167}
  {arXiv:cond-mat/0207167 [cond-mat.str-el]} \BibitemShut {NoStop}%
\bibitem [{\citenamefont {{Nussinov}}\ and\ \citenamefont
  {{Zaanen}}(2002)}]{Zaanen02A}%
  \BibitemOpen
  \bibfield  {author} {\bibinfo {author} {\bibfnamefont {Z.}~\bibnamefont
  {{Nussinov}}}\ and\ \bibinfo {author} {\bibfnamefont {J.}~\bibnamefont
  {{Zaanen}}},\ }\bibfield  {title} {\enquote {\bibinfo {title} {{Stripe
  fractionalization I: the generation of Ising local symmetry}},}\ }\href
  {\doibase 10.1051/jp4:20020405} {\bibfield  {journal} {\bibinfo  {journal}
  {J. Phys. IV France}\ }\textbf {\bibinfo {volume} {12}},\ \bibinfo {pages}
  {245} (\bibinfo {year} {2002})},\ \Eprint
  {http://arxiv.org/abs/cond-mat/0209437} {cond-mat/0209437} \BibitemShut
  {NoStop}%
\bibitem [{\citenamefont {{Zaanen}}\ and\ \citenamefont
  {{Nussinov}}(2003)}]{Zaanen02B}%
  \BibitemOpen
  \bibfield  {author} {\bibinfo {author} {\bibfnamefont {J.}~\bibnamefont
  {{Zaanen}}}\ and\ \bibinfo {author} {\bibfnamefont {Z.}~\bibnamefont
  {{Nussinov}}},\ }\bibfield  {title} {\enquote {\bibinfo {title} {{Stripe
  fractionalization: the quantum spin nematic and the Abrikosov lattice}},}\
  }\href {\doibase 10.1002/pssb.200301673} {\bibfield  {journal} {\bibinfo
  {journal} {Phys. Stat. Sol. B}\ }\textbf {\bibinfo {volume} {236}},\ \bibinfo
  {pages} {332} (\bibinfo {year} {2003})},\ \Eprint
  {http://arxiv.org/abs/cond-mat/0209441} {cond-mat/0209441} \BibitemShut
  {NoStop}%
\bibitem [{\citenamefont {{Grover}}\ and\ \citenamefont
  {{Senthil}}(2010)}]{TGTS10}%
  \BibitemOpen
  \bibfield  {author} {\bibinfo {author} {\bibfnamefont {T.}~\bibnamefont
  {{Grover}}}\ and\ \bibinfo {author} {\bibfnamefont {T.}~\bibnamefont
  {{Senthil}}},\ }\bibfield  {title} {\enquote {\bibinfo {title} {{Quantum
  phase transition from an antiferromagnet to a spin liquid in a metal}},}\
  }\href {\doibase 10.1103/PhysRevB.81.205102} {\bibfield  {journal} {\bibinfo
  {journal} {\prb}\ }\textbf {\bibinfo {volume} {81}},\ \bibinfo {eid} {205102}
  (\bibinfo {year} {2010})},\ \Eprint {http://arxiv.org/abs/0910.1277}
  {arXiv:0910.1277 [cond-mat.str-el]} \BibitemShut {NoStop}%
\bibitem [{\citenamefont {{Kaul}}\ \emph {et~al.}(2008)\citenamefont {{Kaul}},
  \citenamefont {{Metlitski}}, \citenamefont {{Sachdev}},\ and\ \citenamefont
  {{Xu}}}]{Kaul08}%
  \BibitemOpen
  \bibfield  {author} {\bibinfo {author} {\bibfnamefont {R.~K.}\ \bibnamefont
  {{Kaul}}}, \bibinfo {author} {\bibfnamefont {M.~A.}\ \bibnamefont
  {{Metlitski}}}, \bibinfo {author} {\bibfnamefont {S.}~\bibnamefont
  {{Sachdev}}}, \ and\ \bibinfo {author} {\bibfnamefont {C.}~\bibnamefont
  {{Xu}}},\ }\bibfield  {title} {\enquote {\bibinfo {title} {{Destruction of
  N{\'e}el order in the cuprates by electron doping}},}\ }\href {\doibase
  10.1103/PhysRevB.78.045110} {\bibfield  {journal} {\bibinfo  {journal} {Phys.
  Rev. B}\ }\textbf {\bibinfo {volume} {78}},\ \bibinfo {eid} {045110}
  (\bibinfo {year} {2008})},\ \Eprint {http://arxiv.org/abs/0804.1794}
  {arXiv:0804.1794 [cond-mat.str-el]} \BibitemShut {NoStop}%
\bibitem [{\citenamefont {{Mross}}\ and\ \citenamefont
  {{Senthil}}(2012{\natexlab{a}})}]{Mross12}%
  \BibitemOpen
  \bibfield  {author} {\bibinfo {author} {\bibfnamefont {D.~F.}\ \bibnamefont
  {{Mross}}}\ and\ \bibinfo {author} {\bibfnamefont {T.}~\bibnamefont
  {{Senthil}}},\ }\bibfield  {title} {\enquote {\bibinfo {title} {{Theory of a
  Continuous Stripe Melting Transition in a Two-Dimensional Metal: A Possible
  Application to Cuprate Superconductors}},}\ }\href {\doibase
  10.1103/PhysRevLett.108.267001} {\bibfield  {journal} {\bibinfo  {journal}
  {Phys. Rev. Lett.}\ }\textbf {\bibinfo {volume} {108}},\ \bibinfo {eid}
  {267001} (\bibinfo {year} {2012}{\natexlab{a}})},\ \Eprint
  {http://arxiv.org/abs/1201.3358} {arXiv:1201.3358 [cond-mat.str-el]}
  \BibitemShut {NoStop}%
\bibitem [{\citenamefont {{Mross}}\ and\ \citenamefont
  {{Senthil}}(2012{\natexlab{b}})}]{Mross12b}%
  \BibitemOpen
  \bibfield  {author} {\bibinfo {author} {\bibfnamefont {D.~F.}\ \bibnamefont
  {{Mross}}}\ and\ \bibinfo {author} {\bibfnamefont {T.}~\bibnamefont
  {{Senthil}}},\ }\bibfield  {title} {\enquote {\bibinfo {title} {{Stripe
  melting and quantum criticality in correlated metals}},}\ }\href {\doibase
  10.1103/PhysRevB.86.115138} {\bibfield  {journal} {\bibinfo  {journal} {Phys.
  Rev. B}\ }\textbf {\bibinfo {volume} {86}},\ \bibinfo {eid} {115138}
  (\bibinfo {year} {2012}{\natexlab{b}})},\ \Eprint
  {http://arxiv.org/abs/1207.1442} {arXiv:1207.1442 [cond-mat.str-el]}
  \BibitemShut {NoStop}%
\bibitem [{\citenamefont {Hertz}(1976)}]{hertz}%
  \BibitemOpen
  \bibfield  {author} {\bibinfo {author} {\bibfnamefont {J.~A.}\ \bibnamefont
  {Hertz}},\ }\bibfield  {title} {\enquote {\bibinfo {title} {{Quantum critical
  phenomena}},}\ }\href {\doibase 10.1103/PhysRevB.14.1165} {\bibfield
  {journal} {\bibinfo  {journal} {Phys. Rev. B}\ }\textbf {\bibinfo {volume}
  {14}},\ \bibinfo {pages} {1165} (\bibinfo {year} {1976})}\BibitemShut
  {NoStop}%
\bibitem [{\citenamefont {{Read}}\ \emph {et~al.}(1995)\citenamefont {{Read}},
  \citenamefont {{Sachdev}},\ and\ \citenamefont {{Ye}}}]{RSY95}%
  \BibitemOpen
  \bibfield  {author} {\bibinfo {author} {\bibfnamefont {N.}~\bibnamefont
  {{Read}}}, \bibinfo {author} {\bibfnamefont {S.}~\bibnamefont {{Sachdev}}}, \
  and\ \bibinfo {author} {\bibfnamefont {J.}~\bibnamefont {{Ye}}},\ }\bibfield
  {title} {\enquote {\bibinfo {title} {{Landau theory of quantum spin glasses
  of rotors and Ising spins}},}\ }\href {\doibase 10.1103/PhysRevB.52.384}
  {\bibfield  {journal} {\bibinfo  {journal} {Phys. Rev. B}\ }\textbf {\bibinfo
  {volume} {52}},\ \bibinfo {pages} {384} (\bibinfo {year} {1995})},\ \Eprint
  {http://arxiv.org/abs/cond-mat/9412032} {arXiv:cond-mat/9412032 [cond-mat]}
  \BibitemShut {NoStop}%
\bibitem [{\citenamefont {{Georges}}\ \emph {et~al.}(2000)\citenamefont
  {{Georges}}, \citenamefont {{Parcollet}},\ and\ \citenamefont
  {{Sachdev}}}]{GPS00}%
  \BibitemOpen
  \bibfield  {author} {\bibinfo {author} {\bibfnamefont {A.}~\bibnamefont
  {{Georges}}}, \bibinfo {author} {\bibfnamefont {O.}~\bibnamefont
  {{Parcollet}}}, \ and\ \bibinfo {author} {\bibfnamefont {S.}~\bibnamefont
  {{Sachdev}}},\ }\bibfield  {title} {\enquote {\bibinfo {title} {{Mean Field
  Theory of a Quantum Heisenberg Spin Glass}},}\ }\href {\doibase
  10.1103/PhysRevLett.85.840} {\bibfield  {journal} {\bibinfo  {journal} {Phys.
  Rev. Lett.}\ }\textbf {\bibinfo {volume} {85}},\ \bibinfo {pages} {840}
  (\bibinfo {year} {2000})},\ \Eprint {http://arxiv.org/abs/cond-mat/9909239}
  {cond-mat/9909239} \BibitemShut {NoStop}%
\bibitem [{\citenamefont {{Georges}}\ \emph {et~al.}(2001)\citenamefont
  {{Georges}}, \citenamefont {{Parcollet}},\ and\ \citenamefont
  {{Sachdev}}}]{GPS01}%
  \BibitemOpen
  \bibfield  {author} {\bibinfo {author} {\bibfnamefont {A.}~\bibnamefont
  {{Georges}}}, \bibinfo {author} {\bibfnamefont {O.}~\bibnamefont
  {{Parcollet}}}, \ and\ \bibinfo {author} {\bibfnamefont {S.}~\bibnamefont
  {{Sachdev}}},\ }\bibfield  {title} {\enquote {\bibinfo {title} {{Quantum
  fluctuations of a nearly critical Heisenberg spin glass}},}\ }\href {\doibase
  10.1103/PhysRevB.63.134406} {\bibfield  {journal} {\bibinfo  {journal} {Phys.
  Rev. B}\ }\textbf {\bibinfo {volume} {63}},\ \bibinfo {eid} {134406}
  (\bibinfo {year} {2001})},\ \Eprint {http://arxiv.org/abs/cond-mat/0009388}
  {cond-mat/0009388} \BibitemShut {NoStop}%
\bibitem [{\citenamefont {Maldacena}\ and\ \citenamefont
  {Stanford}(2016)}]{Maldacena2016}%
  \BibitemOpen
  \bibfield  {author} {\bibinfo {author} {\bibfnamefont {J.}~\bibnamefont
  {Maldacena}}\ and\ \bibinfo {author} {\bibfnamefont {D.}~\bibnamefont
  {Stanford}},\ }\bibfield  {title} {\enquote {\bibinfo {title} {Remarks on the
  {S}achdev-{Y}e-{K}itaev model},}\ }\href {\doibase
  10.1103/PhysRevD.94.106002} {\bibfield  {journal} {\bibinfo  {journal} {Phys.
  Rev. D}\ }\textbf {\bibinfo {volume} {94}},\ \bibinfo {pages} {106002}
  (\bibinfo {year} {2016})},\ \Eprint {http://arxiv.org/abs/1604.07818}
  {arXiv:1604.07818 [hep-th]} \BibitemShut {NoStop}%
\bibitem [{\citenamefont {Kitaev}\ and\ \citenamefont {Suh}(2018)}]{KitaevSuh}%
  \BibitemOpen
  \bibfield  {author} {\bibinfo {author} {\bibfnamefont {A.}~\bibnamefont
  {Kitaev}}\ and\ \bibinfo {author} {\bibfnamefont {S.~J.}\ \bibnamefont
  {Suh}},\ }\bibfield  {title} {\enquote {\bibinfo {title} {{The soft mode in
  the Sachdev-Ye-Kitaev model and its gravity dual}},}\ }\href {\doibase
  10.1007/JHEP05(2018)183} {\bibfield  {journal} {\bibinfo  {journal} {JHEP}\
  }\textbf {\bibinfo {volume} {05}},\ \bibinfo {pages} {183} (\bibinfo {year}
  {2018})},\ \Eprint {http://arxiv.org/abs/1711.08467} {arXiv:1711.08467
  [hep-th]} \BibitemShut {NoStop}%
\bibitem [{\citenamefont {Gu}\ \emph {et~al.}(2020)\citenamefont {Gu},
  \citenamefont {Kitaev}, \citenamefont {Sachdev},\ and\ \citenamefont
  {Tarnopolsky}}]{Gu2019}%
  \BibitemOpen
  \bibfield  {author} {\bibinfo {author} {\bibfnamefont {Y.}~\bibnamefont
  {Gu}}, \bibinfo {author} {\bibfnamefont {A.}~\bibnamefont {Kitaev}}, \bibinfo
  {author} {\bibfnamefont {S.}~\bibnamefont {Sachdev}}, \ and\ \bibinfo
  {author} {\bibfnamefont {G.}~\bibnamefont {Tarnopolsky}},\ }\bibfield
  {title} {\enquote {\bibinfo {title} {{Notes on the complex Sachdev-Ye-Kitaev
  model}},}\ }\href {\doibase 10.1007/JHEP02(2020)157} {\bibfield  {journal}
  {\bibinfo  {journal} {JHEP}\ }\textbf {\bibinfo {volume} {02}},\ \bibinfo
  {pages} {157} (\bibinfo {year} {2020})},\ \Eprint
  {http://arxiv.org/abs/1910.14099} {arXiv:1910.14099 [hep-th]} \BibitemShut
  {NoStop}%
\bibitem [{\citenamefont {Chubukov}\ \emph {et~al.}(1994)\citenamefont
  {Chubukov}, \citenamefont {Sachdev},\ and\ \citenamefont {Ye}}]{CSY94}%
  \BibitemOpen
  \bibfield  {author} {\bibinfo {author} {\bibfnamefont {A.~V.}\ \bibnamefont
  {Chubukov}}, \bibinfo {author} {\bibfnamefont {S.}~\bibnamefont {Sachdev}}, \
  and\ \bibinfo {author} {\bibfnamefont {J.}~\bibnamefont {Ye}},\ }\bibfield
  {title} {\enquote {\bibinfo {title} {Theory of two-dimensional quantum
  heisenberg antiferromagnets with a nearly critical ground state},}\ }\href
  {\doibase 10.1103/PhysRevB.49.11919} {\bibfield  {journal} {\bibinfo
  {journal} {Phys. Rev. B}\ }\textbf {\bibinfo {volume} {49}},\ \bibinfo
  {pages} {11919} (\bibinfo {year} {1994})}\BibitemShut {NoStop}%
\bibitem [{\citenamefont {{Podolsky}}\ and\ \citenamefont
  {{Sachdev}}(2012)}]{DPSS}%
  \BibitemOpen
  \bibfield  {author} {\bibinfo {author} {\bibfnamefont {D.}~\bibnamefont
  {{Podolsky}}}\ and\ \bibinfo {author} {\bibfnamefont {S.}~\bibnamefont
  {{Sachdev}}},\ }\bibfield  {title} {\enquote {\bibinfo {title} {{Spectral
  functions of the Higgs mode near two-dimensional quantum critical points}},}\
  }\href {\doibase 10.1103/PhysRevB.86.054508} {\bibfield  {journal} {\bibinfo
  {journal} {Phys. Rev. B}\ }\textbf {\bibinfo {volume} {86}},\ \bibinfo {eid}
  {054508} (\bibinfo {year} {2012})},\ \Eprint {http://arxiv.org/abs/1205.2700}
  {arXiv:1205.2700 [cond-mat.quant-gas]} \BibitemShut {NoStop}%
\bibitem [{\citenamefont {Amit}(1984)}]{amit1984field}%
  \BibitemOpen
  \bibfield  {author} {\bibinfo {author} {\bibfnamefont {D.}~\bibnamefont
  {Amit}},\ }\href {https://books.google.com/books?id=M4yqQgAACAAJ} {\emph
  {\bibinfo {title} {{Field Theory, the Renormalization Group, and Critical
  Phenomena}}}},\ {International Series in Pure and Applied Physics}\ (\bibinfo
   {publisher} {World Scientific},\ \bibinfo {year} {1984})\BibitemShut
  {NoStop}%
\bibitem [{\citenamefont {{Jian}}\ \emph {et~al.}(2020)\citenamefont {{Jian}},
  \citenamefont {{Xu}}, \citenamefont {{Wu}},\ and\ \citenamefont
  {{Xu}}}]{XuChao-Ming}%
  \BibitemOpen
  \bibfield  {author} {\bibinfo {author} {\bibfnamefont {C.-M.}\ \bibnamefont
  {{Jian}}}, \bibinfo {author} {\bibfnamefont {Y.}~\bibnamefont {{Xu}}},
  \bibinfo {author} {\bibfnamefont {X.-C.}\ \bibnamefont {{Wu}}}, \ and\
  \bibinfo {author} {\bibfnamefont {C.}~\bibnamefont {{Xu}}},\ }\bibfield
  {title} {\enquote {\bibinfo {title} {{Continuous N{\'e}el-VBS Quantum Phase
  Transition in Non-Local one-dimensional systems with SO(3) Symmetry}},}\
  }\href@noop {} {\  (\bibinfo {year} {2020})},\ \Eprint
  {http://arxiv.org/abs/2004.07852} {arXiv:2004.07852 [cond-mat.str-el]}
  \BibitemShut {NoStop}%
\end{thebibliography}

%

\end{document}